\documentclass[paper,11pt]{JHEP}
\usepackage[centertags]{amsmath}
\usepackage{amsfonts} \usepackage{amssymb} \usepackage{amsthm}
\allowdisplaybreaks[1]
\usepackage{graphicx}
\newcommand{\Z}{{\mathbb Z}}

\newcommand{\ii}{\mathrm{i}}
\newcommand{\ee}{\mathrm{e}}

\newcommand{\cF}{{\mathcal{F}}}

\newcommand{\cN}{{\mathcal{N}}}
\newcommand{\cO}{{\mathcal{O}}}

\newcommand{\cS}{{\mathcal{S}}}
\newcommand{\cT}{{\mathcal{T}}}

\newcommand{\cW}{{\mathcal{W}}}

\newcommand{\one}{{\rm 1\kern -.9mm l}}

\newcommand{\spin}[2]{\big[^{#1}_{#2}\big]}
\newcommand{\Pf}{\mathrm{Pf}}

\newcommand{\pffb}{\Big(\frac{\partial~}{\partial\varphi_2}\Big)}

\newcommand{\bin}[2]{\Big( {\begin{array}{*{20}c} #1 \\ #2 \\ \end{array}} \Big)}

\newdimen\tableauside\tableauside=1.0ex
\newdimen\tableaurule\tableaurule=0.4pt
\newdimen\tableaustep
\def\phantomhrule#1{\hbox{\vbox to0pt{\hrule height\tableaurule
width#1\vss}}}
\def\phantomvrule#1{\vbox{\hbox to0pt{\vrule width\tableaurule
height#1\hss}}}
\def\sqr{\vbox{%
  \phantomhrule\tableaustep
\hbox{\phantomvrule\tableaustep\kern\tableaustep\phantomvrule\tableaustep}%
  \hbox{\vbox{\phantomhrule\tableauside}\kern-\tableaurule}}}
\def\squares#1{\hbox{\count0=#1\noindent\loop\sqr
  \advance\count0 by-1 \ifnum\count0>0\repeat}}
\def\tableau#1{\vcenter{\offinterlineskip
  \tableaustep=\tableauside\advance\tableaustep by-\tableaurule
  \kern\normallineskip\hbox
    {\kern\normallineskip\vbox
      {\gettableau#1 0 }%
     \kern\normallineskip\kern\tableaurule}%
  \kern\normallineskip\kern\tableaurule}}
\def\gettableau#1 {\ifnum#1=0\let\next=\null\else
  \squares{#1}\let\next=\gettableau\fi\next}
\newcommand{\Yfund}{\tableau{1}}
\newcommand{\Ysymm}{\tableau{2}}
\newcommand{\Yasymm}{\tableau{1 1}}
\newcommand{\eu}{\epsilon_1}
\newcommand{\ed}{\epsilon_2}
\newcommand{\Smu}{\sum_f m_f}
\newcommand{\Smumu}{\sum_{f<f'} m_f m_{f'}}

\title{Deformed $\mathcal N=2$ theories, generalized recursion relations and S-duality
}
\author{M. Bill\'o$^1$, M. Frau$^{1}$, L. Gallot$^{2}$, A. Lerda$^{3}$, I. Pesando$^{1}$
\\
\vskip 0.2cm
$^1$ Universit\`a di Torino, Dipartimento di Fisica
and I.N.F.N. - sezione di Torino \\
Via P. Giuria 1, I-10125 Torino, Italy\\
\vskip 0.2cm
$^2$LAPTH, Universit\'e de Savoie, CNRS\\
9, Chemin de Bellevue,
74941 Annecy le Vieux Cedex, France\\
\vskip 0.2cm
$^3$Universit\`a del Piemonte Orientale, Dipartimento di Scienze e Innovazione Tecnologica, \\
and I.N.F.N. - Gruppo Collegato di Alessandria - sezione di Torino\\
Viale T. Michel  11, I-15121 Alessandria, Italy\\
\vspace{0.25cm}
\email{billo,frau,lerda,ipesando@to.infn.it; laurent.gallot@lapp.in2p3.fr} 
}
\abstract{We study the non-perturbative properties of $\cN=2$ super conformal
field theories in four dimensions using localization techniques. In particular we consider
SU(2) gauge theories, deformed by a generic $\epsilon$-background, with four fundamental flavors or with one adjoint hypermultiplet. 
In both cases we explicitly compute
the first few instanton corrections to the partition  function and the
prepotential using Nekrasov's approach.
These results allow to reconstruct exact expressions
involving quasi-modular functions of the bare gauge
coupling constant and to show that the prepotential terms
satisfy a modular anomaly equation that takes the form of a recursion relation with an explicitly
$\epsilon$-dependent term. We then investigate the implications of this recursion relation on the 
modular properties of the effective theory and find
that with a suitable redefinition of the prepotential and of the
effective coupling it is possible, at least up to the third order in the deformation parameters, 
to cast the S-duality relations in the same form as they appear in the Seiberg-Witten solution of the undeformed theory.}

\preprint{
LAPTH 005/13}

\begin{document}

\section{Introduction}
\label{secn:intro}
Four-dimensional field theories with rigid $\cN=2$ supersymmetry provide a remarkable arena where
many exact results can be obtained; indeed $\cN=2$ supersymmetry, not being maximal,
allows for a great deal of flexibility but, at the same time, is large enough to guarantee 
full control. This fact was exploited in 
the seminal papers \cite{Seiberg:1994rs,Seiberg:1994aj} where it was shown that
the effective dynamics of $\cN=2$ super Yang-Mills (SYM) theories
in the limit of low energy and momenta can be
exactly encoded in the so-called Seiberg-Witten (SW) curve describing the geometry of the moduli
space of the SYM vacua. When the gauge group is SU(2), 
the SW curve defines a torus whose complex structure parameter is 
identified with the (complexified) gauge coupling constant $\tau$ of the SYM theory at low energy. 
This coupling receives perturbative corrections at 1-loop and non-perturbative corrections due to instantons, and the corresponding effective action follows from a prepotential $\cF$ that is a holomorphic function of the vacuum expectation value $a$ of the adjoint vector multiplet, 
of the flavor masses, if any, and of the dynamically generated scale in asymptotically free theories
or of the bare gauge coupling constant $\tau_0$ in conformal models
(see for instance \cite{D'Hoker:1999ft} for a review and extensions of this approach).

Recently, $\cN=2$ superconformal field theories (SCFT) have attracted a lot 
of attention. Two canonical examples of SCFT's
are the $\cN=2$ SU(2) SYM theory with $N_f=4$ fundamental hypermultiplets
and the $\cN=2^*$ theory, namely a $\cN=2$ SYM theory with an adjoint hypermultiplet. 
In both cases, the $\beta$-function vanishes but, when the 
hypermultiplets are massive, the bare coupling $\tau_0$ gets renormalized at 1-loop 
by terms proportional to the mass parameters. 
Besides these, there are also non-perturbative corrections due to instantons.
As shown in \cite{Minahan:1997if} for the $\cN=2^*$ theory and more recently in \cite{Billo':2011pr} for the $N_f=4$ theory, by organizing the effective prepotential $\cF$ as a series 
in inverse powers of $a$ and by exploiting a recursion relation hidden in the SW curve, 
it is possible to write the various terms of $\cF$ 
as \emph{exact} functions of the bare coupling.
These functions are polynomials in Eisenstein series and Jacobi $\theta$-functions
of $\tau_0$ and their modular properties allow to show that the effective theory at low energy
inherits the $\mathrm{Sl}(2,\mathbb{Z})$ symmetry of the microscopic theory at high energy.
In particular one can show \cite{Minahan:1997if,Billo':2011pr} that the S-duality map on the
bare coupling, {\it i.e.} $\tau_0\to-{1}/{\tau_0}$, implies the corresponding map on the effective
coupling, {\it i.e.} $\tau\to-{1}/{\tau}$, and that the prepotential 
$\cF$ and its S-dual are related to each other by a Legendre transformation.

The non-perturbative corrections predicted by the SW solution 
can also be obtained directly via multi-instanton calculus and the use of localization techniques \cite{Nekrasov:2002qd,Nekrasov:2003rj}%
\footnote{See also \cite{Losev:1997bz,Moore:1998et} for earlier applications of these techniques.}. This approach is based on the calculation of the instanton
partition function after introducing two deformation parameters, $\eu$ and $\ed$, 
of mass dimension 1 
which break the four-dimensional Lorentz invariance, regularize the space-time volume 
and fully localize the integrals over the instanton moduli space on sets of
isolated points, thus allowing their explicit evaluation. This method, which has been extensively applied to many models (see for instance 
\cite{Flume:2002az}\,-\,\nocite{Bruzzo:2002xf,Flume:2004rp,Nekrasov:2004vw,Marino:2004cn,Billo:2009di,Fucito:2009rs}\cite{Billo':2010bd}) can be interpreted as the effect of putting the gauge theory in a curved background, known 
as $\Omega$-background \cite{Nekrasov:2002qd,Nekrasov:2003rj,Nakajima:2003uh}, 
or in a supergravity background with a non-trivial graviphoton field strength, 
which are equivalent on the instanton moduli space \cite{Billo:2006jm,Ito:2010vx}. 
The resulting instanton partition function $Z_{\mathrm{inst}}(\eu,\ed)$, 
also known as Nekrasov partition function, allows to obtain the non-perturbative part of
the SYM prepotential according to
\begin{equation}
\cF_{\mathrm{inst}}= -\lim_{\eu,\ed\to 0}\epsilon_1\epsilon_2\,\log Z_{\mathrm{inst}}(\eu,\ed)~.
\label{FlogZo}
\end{equation}
Actually, the Nekrasov partition function is useful not only when the $\epsilon$ parameters are sent
to zero, but also when they are kept at finite values. In this case, in fact, the non-perturbative $\epsilon$-deformed prepotential
\begin{equation}
F_{\mathrm{inst}}(\eu,\ed)= -\epsilon_1\epsilon_2\,\log Z_{\mathrm{inst}}(\eu,\ed)
\label{FlogZ}
\end{equation}
represents a very interesting generalization of the SYM one. By adding to it the 
corresponding ($\epsilon$-deformed) perturbative part $F_{\mathrm{pert}}$, one gets a 
generalized prepotential that can be conveniently expanded as follows
\begin{equation}
 F_{\mathrm{pert}}+F_{\mathrm{inst}}
=\sum_{n,g=0}^\infty F^{(n,g)}\,(\epsilon_1+\epsilon_2)^{2n}\,(\epsilon_1\epsilon_2)^g~.
\label{Fng}
\end{equation}
The amplitude $F^{(0,0)}$, which is the only one that remains
when the $\epsilon$-deformations are switched off, 
coincides with the SYM prepotential $\cF$ of the SW theory, 
up to the classical tree-level term.
The amplitudes $F^{(0,g)}$ with $g\geq 1$ 
account instead for gravitational couplings and correspond to F-terms in the effective action of the 
form $F^{(0,g)}\cW^{2g}$, 
where $\cW$ is the chiral Weyl superfield containing the graviphoton field strength as its lowest component. These terms were obtained long ago from the genus $g$ partition function of the $\cN=2$
topological string on an appropriate Calabi-Yau background \cite{Antoniadis:1993ze} and were shown to satisfy a holomorphic anomaly equation \cite{Bershadsky:1993ta,Bershadsky:1993cx} which allows to recursively reconstruct
the higher genus contributions from the lower genus ones (see for instance 
\cite{Klemm:2002pa,Huang:2009md}).
More recently, also the amplitudes $F^{(n,g)}$ 
with $n\not=0$ have been related to the $\cN=2$ topological string and have been 
shown to correspond to higher dimensional F-terms of the type $F^{(n,g)}\varUpsilon^{2n}\cW^{2g}$
where $\varUpsilon$ is a chiral projection of real functions of $\cN=2$ vector superfields \cite{Antoniadis:2010iq}, which also satisfy an extended holomorphic anomaly equation \cite{Krefl:2010fm,Huang:2010kf}. By taking the limit $\ed\to 0$ with $\eu$ finite, one selects 
in (\ref{Fng}) the amplitudes $F^{(n,0)}$. This limit, also known as Nekrasov-Shatashvili limit \cite{Nekrasov:2009rc}, is particularly interesting since it is believed that
the $\cN=2$ effective theory can be described in this case by certain quantum integrable systems.
Furthermore, in this Nekrasov-Shatashvili limit, using saddle point methods it is possible to
derive a generalized SW curve \cite{Poghossian:2010pn,Fucito:2011pn} and extend the 
above-mentioned results for the SYM theories to the $\epsilon$-deformed ones.

By considering the Nekrasov partition function and the corresponding generalized prepotential 
for rank one SCFT's, in \cite{Alday:2009aq} a very remarkable relation has been uncovered
with the correlation functions of a two-dimensional Liouville theory with an $\epsilon$-dependent
central charge. 
In particular, for the SU(2) theory with $N_f=4$ the generalized
prepotential turns out to be related to the logarithm of the conformal blocks of four 
Liouville operators on a sphere and the
bare gauge coupling constant to the cross-ratio of the punctures where the four operators are located; 
for the $\cN=2^*$ SU(2) theory, instead,  the correspondence works with the one-point conformal 
blocks on a torus whose complex structure parameter plays the role of the complexified
bare gauge coupling. Since the conformal blocks of the Liouville theory have 
well-defined properties under modular transformations, it is natural to explore modularity 
also on the four-dimensional gauge theory and try to connect it to 
the strong/weak-coupling S-duality,
generalizing in this way the SW results when
$\eu$ and $\ed$ are non-zero. On the other hand, one expects that the 
deformed gauge theory should somehow inherit the duality properties of the Type IIB string 
theory in which it can be embedded.
Some important steps towards this goal have been 
made in \cite{Gaiotto:2009we} and also in \cite{Mironov:2009ib} where it has been shown that
the SW contour integral techniques remain valid also when both $\eu$ and $\ed$ are
non-vanishing. To make further progress and gain a more quantitative understanding, it would be 
useful to know the various amplitudes $F^{(n,g)}$ in (\ref{Fng}) as \emph{exact} functions of the gauge coupling constant and analyze 
their behavior under modular transformations, similarly to what has
been done for the SW prepotential $F^{(0,0)}$ in \cite{Minahan:1997if,Billo':2011pr}. 
Recently, by exploiting the generalized holomorphic anomaly equation, an exact expression
in terms of Eisenstein series has been given for the first few amplitudes $F^{(n,g)}$ of the SU(2) 
theory with $N_f=4$ and the $\cN=2^*$ SU(2) theory in the limit of vanishing hypermultiplet masses
\cite{Huang:2011qx}. This analysis has then been extended in \cite{Huang:2012kn}
to the massive $\cN=2^*$ model in the Nekrasov-Shatashvili limit, 
using again the extended holomorphic anomaly equation, and in \cite{KashaniPoor:2012wb} 
using the properties of the Liouville toroidal conformal blocks in the semi-classical limit of infinite 
central charge. 
However, finding the modular properties of the deformed prepotential in full generality
still remains an open issue.

In this paper we address this problem and extend the previous results
by adopting a different strategy. In Section~\ref{secn:N2star}, using localization techniques 
we explicitly compute the first few instanton corrections to the prepotential for the $\cN=2^*$ massive
theory with gauge group SU(2) in a generic $\epsilon$-background. {From} these explicit
results we then infer the exact expressions of the various prepotential coefficients and 
write them in terms of Eisenstein series of the bare coupling. Our results reduce to those of \cite{Minahan:1997if,Billo':2011pr} when the deformation parameters are switched off, and to
those of \cite{Huang:2011qx}\,-\,\nocite{Huang:2012kn}\cite{KashaniPoor:2012wb} 
in the massless or in the Nekrasov-Shatashvili limits. The properties of the Eisenstein series
allow to analyze the behavior of the various prepotential terms under modular transformations
and also to write a recursion relation that is equivalent to the holomorphic anomaly equation if one trades modularity for holomorphicity. The recursion relation we find contains a
term proportional to $\eu\ed$, which is
invisible in the SYM limit or in the Nekrasov-Shatashvili limit.
In Section~\ref{secn:nf4}
we repeat the same steps for the $\cN=2$ SU(2) theory with $N_f=4$ and arbitrary mass parameters
and also in this case derive the modular anomaly equation in the form of a recursion relation.
In Section~\ref{secn:sdual} we study in detail the properties of the generalized prepotential
under S-duality, and show that when both $\eu$ and $\ed$ are different from zero, due to the 
new term in the recursion relation, the prepotential
and its S-dual are not any more related by a Legendre transformation, an observation which has been
recently put forward in \cite{Galakhov:2012gw} from a different perspective. We also propose how
the relation between the prepotential and its S-dual has to be modified, by computing the first
corrections in $\eu\ed$.
Finally, in Section~\ref{secn:conl} we conclude by showing that there exist
suitable redefinitions of the prepotential and of the effective coupling that allow to recover
the simple Legendre relation and write the S-duality relations in the same form as in the undeformed
SW theory. 
The appendices contain some technical details and present several explicit formulas which are useful for the computations described in the main text.

\section{The $\cN=2^*$ SU(2) theory}
\label{secn:N2star}
The $\cN=2^*$ SYM theory describes the interactions of a $\cN=2$ gauge vector multiplet 
with a massive $\cN=2$ hypermultiplet in the adjoint representation of the gauge group.
It can be regarded as a massive deformation of the $\cN=4$ SYM theory in which the $\beta$-function
remains vanishing but the gauge coupling constant receives both perturbative and non-perturbative
corrections proportional to the hypermultiplet mass. Using the localization techniques 
\cite{Nekrasov:2002qd,Nekrasov:2003rj} one can obtain a generalization of this theory
by considering the $\epsilon$-dependent terms in the Nekrasov partition function.
In the following we only discuss the case in which the gauge group is 
SU(2) (broken down to U(1) by the vacuum expectation value $a$ of the adjoint scalar of the
gauge vector multiplet). 
We begin by considering the non-perturbative corrections.

\subsection{Instanton partition functions}
\label{subsecn:inst2*}
The partition function $Z_k$ at instanton number $k$ is defined by the following
integral over the instanton moduli space ${\mathcal M_k}$:
\begin{equation}
 Z_k=\int d{\mathcal M_k} \,\ee^{-S_{\mathrm{inst}}}
\label{Zk}
\end{equation}
where $S_{\mathrm{inst}}$ is the instanton moduli action of the $\cN=2^*$ theory. 
After introducing two deformation
parameters $\epsilon_1$ and $\epsilon_2$, the partition function $Z_k$ can be explicitly computed using the localization techniques. 
In the case at hand, each $Z_k$ can be expressed
as a sum of terms in one-to-one correspondence with an ordered pair of Young tableaux of U($k$) such that the total number of boxes in the two tableaux is $k$. For example, at $k=1$ we have 
the two possibilities: $(\Yfund,\bullet)$ and $(\bullet,\Yfund)$; at $k=2$ we have instead 
the five cases: $(\Ysymm,\bullet)$, $(\bullet,\Ysymm)$, $(\Yasymm,\bullet)$, $(\bullet,\Yasymm)$, $(\Yfund,\Yfund)$; and so on and so forth. 
Referring for example to the Appendix A of \cite{Billo:2012st} for details, 
at $k=1$ one finds
\begin{equation}
 \begin{aligned}
  Z_{(\Yfund,\bullet)}&=\frac{(-\epsilon_1+\widetilde m)(-\epsilon_2+\widetilde m) (a_{12}+\widetilde m)
 (a_{21}+\widetilde m-\epsilon_1-\epsilon_2)}{(-\epsilon_1)
(-\epsilon_2) a_{12} (a_{21}-\epsilon_1-\epsilon_2)}~,\\
Z_{(\bullet,\Yfund)}&=\frac{(-\epsilon_1+\widetilde m)(-\epsilon_2+\widetilde m) 
 (a_{21}+\widetilde m)(a_{12}+\widetilde m-\epsilon_1-\epsilon_2)}{(-\epsilon_1)
(-\epsilon_2)a_{21}(a_{12}-\epsilon_1-\epsilon_2)}~,
 \end{aligned}
\label{Z1}
\end{equation}
where $a_{uv}=a_u-a_v$ with $a_1=-a_2=a$, and
\begin{equation}
 \widetilde m = m+\frac{\epsilon_1+\epsilon_2}{2}
\label{tildem}
\end{equation}
is the equivariant mass parameter for the adjoint hypermultiplet in the $\epsilon$-background
\cite{Okuda:2010ke,Huang:2011qx}.
The 1-instanton partition function is therefore
\begin{equation}
 Z_1= Z_{(\Yfund,\bullet)}+Z_{(\bullet,\Yfund)} 
=\frac{(4m^2-(\epsilon_1-\epsilon_2)^2)(16a^2-4m^2-3(\epsilon_1+\epsilon_2)^2)}{8\epsilon_1
\epsilon_2(4a^2-(\epsilon_1+\epsilon_2)^2)}~.
\label{Z1tot}
\end{equation}
At $k=2$ the relevant partition functions are
\begin{equation}
 \begin{aligned}
Z_{(\Ysymm,\bullet)}&= \frac{(-\epsilon_1+\widetilde m)
(-\epsilon_2+\widetilde m)(-\epsilon_1+\epsilon_2+\widetilde m)(-2\epsilon_2+\widetilde m)(a_{12}+\widetilde m)}
{(-\epsilon_1)(-\epsilon_2)(-\epsilon_1+\epsilon_2)(-2\epsilon_2)a_{12}}
\\&~~~~~\times~\frac{(a_{12}+\widetilde m+\epsilon_2)(a_{21}+\widetilde m-\epsilon_1-\epsilon_2)(a_{21}+\widetilde m-\epsilon_1-2\epsilon_2)}
{(a_{12}+\epsilon_2)(a_{21}-\epsilon_1-\epsilon_2)(a_{21}-\epsilon_1-2\epsilon_2)}
\phantom{\Bigg|}~,\\
Z_{(\Yfund,\Yfund)}&=\frac{(-\epsilon_1+\widetilde m)^2 (-\epsilon_2+\widetilde m)^2 
(a_{12}+\widetilde m-\epsilon_1)
(a_{12}+\widetilde m-\epsilon_2)}{(-\epsilon_1)^2 (-\epsilon_2)^2 (a_{12}-\epsilon_1)
(a_{12}-\epsilon_2)}\phantom{\Bigg|}
\\&~~~~~\times~
\frac{(a_{21}+\widetilde m-\epsilon_1)(a_{21}+\widetilde m-\epsilon_2)}
{(a_{21}-\epsilon_1)(a_{21}-\epsilon_2)}
~.
 \end{aligned}
\label{Z2}
\end{equation}
The contributions corresponding to the other Young tableaux at $k=2$ can be obtained from the previous expressions with suitable redefinitions. In particular, $Z_{(\bullet,\Ysymm)}$ 
is obtained from $Z_{(\Ysymm,\bullet)}$ in (\ref{Z2}) by exchanging $a_{12}\leftrightarrow a_{21}$,
$Z_{(\Yasymm,\bullet)}$ is obtained by exchanging $\epsilon_1\leftrightarrow\epsilon_2$, and finally
$Z_{(\bullet,\Yasymm)}$ is obtained by simultaneously exchanging $a_{12}\leftrightarrow a_{21}$
and $\epsilon_1\leftrightarrow\epsilon_2$.
The complete 2-instanton partition function is
\begin{equation}
Z_2 = Z_{(\Ysymm,\bullet)}+Z_{(\bullet,\Ysymm)}+Z_{(\Yasymm,\bullet)}+Z_{(\bullet,\Yasymm)}
+Z_{(\Yfund,\Yfund)}
 \label{Z2tot}
\end{equation}
but we refrain from writing its expression since it is not particularly inspiring.
This procedure can be systematically extended to higher instanton numbers leading to explicit
formulas for the instanton partition functions.

Following \cite{Nekrasov:2003rj} we can cast these results in a nice and compact form. 
Indeed, defining
\begin{equation}
 q=\ee^{\pi\ii\tau_0} 
\label{q}
\end{equation}
where $\tau_0=\frac{\theta}{2\pi}+\ii\,\frac{4\pi}{g^2}$ is the complexified
gauge coupling constant of the $\cN=2^*$ SYM theory, the grand-canonical instanton
partition function
\begin{equation}
 Z_{\mathrm{inst}}=\sum_{k=0}^\infty {q^{2k}}\,Z_k
\label{zinst}
\end{equation}
where $Z_0=1$, can be rewritten as
\begin{equation}
\begin{aligned}
 Z_{\mathrm{inst}} = \sum_{(Y_1,Y_2)} q^{2|Y|}\,\prod_{i,j=1}^\infty\prod_{u,v=1}^2&
\Bigg[\frac{a_{uv}+\epsilon_1(i-1)-\epsilon_2j}{a_{uv}+\epsilon_1(i-1-\widetilde k_{vj})-\epsilon_2(j-k_{ui})}\\
&
~~\times ~\frac{a_{uv}+\widetilde m+\epsilon_1(i-1-\widetilde k_{vj})-\epsilon_2(j-k_{ui})}{a_{uv}+\widetilde m+\epsilon_1(i-1)-\epsilon_2j}\Bigg]
\end{aligned}
\label{zno}
\end{equation}
where the first line represents the contribution of the gauge vector multiplet and the second line that
of the adjoint hypermultiplet.
Here $k_{ui}$ and $\widetilde k_{ui}$ denote, respectively, the number of boxes in the $i$-th row 
and in the $i$-th column of a Young tableaux $Y_u$  and are related to the Dynkin indices
of the corresponding representation. These quantities can be extended for any integer $i$ with the convention that $k_{ui}=0$ or $\widetilde k_{ui}=0$ if the $i$-th row or the $i$-th column of 
$Y_u$ is empty. For example, for $Y_u=\Ysymm$ the $k_{ui}$'s are $(2,0,0,0,\ldots)$ and the
$\widetilde k_{ui}$'s are $(1,1,0,0,\ldots)$. Moreover we have 
\begin{equation}
|Y|= \sum_{u,i}k_{ui}=\sum_{u,i}\widetilde k_{ui} = k~.
\end{equation}
It is quite straightforward to check that the expressions (\ref{Z1}) and (\ref{Z2}) are
reproduced by the compact formula (\ref{zno}) by selecting the appropriate Young tableaux.

Following Nekrasov's prescription, we can obtain the generalized non-perturbative prepotential 
according to
\begin{equation}
 F_{\mathrm{inst}} = -\epsilon_1 \epsilon_2 \,\log Z_{\mathrm{inst}} = \sum_{k=1}^\infty
q^{2k}\,F_k~.
\label{finst}
\end{equation}
In the limit $\epsilon_\ell\to 0$, the above expression computes the instanton
contributions to the prepotential of the $\cN=2^*$ SYM theory, while the finite $\epsilon$-dependent
terms represent further non-perturbative corrections. Notice that in the limit $\widetilde m\to 0$,
(\ref{zno}) correctly reduces to the partition function of the $\cN=4$ SYM theory, namely to the Euler 
characteristic of the instanton moduli space (see for instance \cite{Fucito:2004ry}).

\subsection{Perturbative part}
The compact expression (\ref{zno}) allows to ``guess'' the perturbative part of the partition
function by applying the same formal reasoning of Section 3.10 of \cite{Nekrasov:2002qd}. Indeed,
in (\ref{zno}) we recognize the following universal ({\it i.e.} $k$-independent but $a$-dependent)
factor 
\begin{equation}
\prod_{i,j=1}^\infty\,\mathop{\prod_{u,v=1}^2}_{u\not=v}
\frac{a_{uv}+\epsilon_1(i-1)-\epsilon_2j}{a_{uv}+\widetilde m+\epsilon_1(i-1)-\epsilon_2j}
\label{Z1loop}
\end{equation}
which, if suitably interpreted and regularized \cite{Nekrasov:2002qd,Nekrasov:2003rj}, can be 
related to the perturbative part of the partition function of the
$\cN=2^*$ theory in the $\epsilon$-background. According to this idea, we are then led 
to write 
\begin{equation}
 F_{\mathrm{pert}} = \epsilon_1\epsilon_2\mathop{\sum_{u,v=1}^2}_{u\not=v}\,\sum_{i,j=1}^\infty 
\log\frac{a_{uv}+\epsilon_1(i-1)-\epsilon_2j}{a_{uv}+\widetilde m+\epsilon_1(i-1)-\epsilon_2j}~.
\label{F1loop}
\end{equation}
Using the following representation for the logarithm
\begin{equation}
 \log\frac{x}{\Lambda}= -\left.\frac{d}{ds}\Big(\frac{\Lambda^s}{\Gamma(s)}
\int_0^\infty \frac{dt}{t} \,t^s\,\ee^{-tx}\Big)\right|_{s=0}
\label{log}
\end{equation}
where $\Lambda$ is an arbitrary scale and summing over $i$ and $j$, we can rewrite (\ref{F1loop}) 
as
\begin{equation}
 F_{\mathrm{pert}} = \epsilon_1\epsilon_2\mathop{\sum_{u,v=1}^2}_{u\not=v}\,\Big[
\gamma_{\epsilon_1,\epsilon_2}(a_{uv}) -\gamma_{\epsilon_1,\epsilon_2}(a_{uv}+\widetilde m)\Big] 
\label{F1loop1}
\end{equation}
where (see also \cite{Nakajima:2003uh,Huang:2011qx})
\begin{equation}
 \gamma_{\epsilon_1,\epsilon_2}(x) = \left.\frac{d}{ds}\Big(\frac{\Lambda^s}{\Gamma(s)}
\int_0^\infty \frac{dt}{t}\frac{t^s\,\ee^{-tx}}{(\ee^{-\epsilon_1t}-1)
(\ee^{-\epsilon_2t}-1)}\Big)\right|_{s=0}~.
\label{gammae1e2}
\end{equation}
This function, which is related to the logarithm of the Barnes double $\Gamma$-function, can be easily computed by expanding for small values of $\epsilon_1$ and $\epsilon_2$.
As a result, $F_{\mathrm{pert}}$ becomes a series in inverse powers of $a^2$ whose first few
terms are
\begin{eqnarray}
F_{\mathrm{pert}} &=&\frac{1}{4}\big(4m^2-s^2\big)\log\frac{2a}{\Lambda}
-\frac{\big(4m^2-s^2\big)\big(4m^2+4p-s^2\big)}{768\,a^2}\nonumber\\
&&\!\!\!-\frac{\big(4m^2-s^2\big)\big(4m^2+4p-s^2\big)\big(4m^2+6p-3s^2\big)}{61440\,a^4}
\nonumber\\
&&\!\!\!-\frac{\big(4m^2-s^2\big)\big(4m^2+4p-s^2\big)\big(48m^4+176m^2p+160p^2-88m^2s^2-204ps^2+51s^4\big)}
{8257536\,a^6}
\nonumber\\&&\!\!\!+\,\cO(a^{-8})\label{F1loopexp}~.
\end{eqnarray}
Here we have used (\ref{tildem}) and introduced the convenient notation
\begin{equation}
 s= \epsilon_1+\epsilon_2~~~,~~~p=\epsilon_1\epsilon_2~.
\label{sp}
\end{equation}
One can easily check that in the limit $\epsilon_\ell\to0$ this expression correctly reproduces the 1-loop prepotential of the $\cN=2^*$ SU(2) gauge theory (see for instance \cite{Minahan:1997if}). 

\subsection{Generalized prepotential}
The complete generalized prepotential is the sum of the classical, perturbative and non-perturbative parts. We now focus on the latter two terms which are directly related to the Nekrasov partition function.
Just like the 1-loop piece (\ref{F1loopexp}), also the instanton terms (\ref{finst}) can be
organized as a series expansion for small values of $\epsilon_1$ and $\epsilon_2$,
or equivalently for large values of $a$. Discarding $a$-independent terms, which are not
relevant for the gauge theory dynamics, we write
\begin{equation}
F_{\mathrm{pert}}+F_{\mathrm{inst}}= h_0 \log\frac{2a}{\Lambda}-\sum_{\ell=1}^\infty \frac{h_\ell}{2^{\ell+1}\,\ell}\,\frac{1}{a^{2\ell}}
\label{fexp}
\end{equation}
where the coefficients $h_\ell$ are polynomials in $m^2$, $s^2$ and $p$ which
can be explicitly derived from (\ref{F1loopexp}) as far as the perturbative part is concerned, and from the instanton partition functions (\ref{zno}), after
using (\ref{finst}), for the non-perturbative part. 
Here we list the first few of these coefficients up to three instantons:
 \begin{align}
  h_0&=\frac{1}{4}\big(4m^2-s^2\big)~,\phantom{\Big|}\label{h0q}\\
h_1&=\big(4m^2-s^2\big)\big(4m^2+4p-s^2\big)\Big(\frac{1}{192}-\frac{1}{8}q^2-\frac{3}{8}q^4
-\frac{1}{2}q^6+\cdots\Big)~,\phantom{\Big|}\label{h1q}\\
h_2&=\big(4m^2-s^2\big)\big(4m^2+4p-s^2\big)\Big(\frac{4m^2+6p-3s^2}{3840}
-\frac{s^2}{8}q^2\nonumber\\\phantom{\Big|}&\hspace{0.7cm}+\frac{12m^2+18p-21s^2}{16}q^4
+\frac{8m^2+12p-9s^2}{2}q^6+\cdots\Big)~,\phantom{\Big|}\label{h2q}\\
h_3&=\big(4m^2-s^2\big)\big(4m^2+4p-s^2\big)\Big(\frac{48m^4+176m^2p+160p^2-88m^2s^2-204ps^2+51s^4}
{172032}\phantom{\Big|}
\nonumber\\&\hspace{0.7cm}
-\frac{3s^4}{32}q^2
-\frac{240m^4+1392m^2p+1440p^2-3000m^2s^2-7548ps^2+3903s^4}{1024}q^4\phantom{\Big|}
\nonumber\\&\hspace{0.7cm}
-\frac{240m^4+1008m^2p+960p^2-1080m^2s^2-2652ps^2+987s^4}{32}q^6+\cdots\Big)\phantom{\Big|}~.\label{h3q}
 \end{align}
It is interesting to notice that all $h_\ell$'s are proportional to $\big(4m^2-s^2\big)$ and,
except for $h_0$, also to $\big(4m^2+4p-s^2\big)$. Explicit expressions for $h_\ell$ with $\ell>3$
can be systematically derived from the generalized prepotential but they are not 
needed for our considerations.

Building on previous results obtained in $\epsilon_\ell\to 0$ limit from the SW curve 
\cite{Minahan:1997if,Billo':2011pr} and on the analysis of \cite{Huang:2011qx} for the massless 
case, we expect that the expressions (\ref{h0q})-(\ref{h3q}) are just the first terms 
in the instanton expansion of (quasi) modular functions of $q$. 
More precisely, we expect that $h_\ell$ are (quasi) modular functions of weight $2\ell$ that can be written solely in terms of the Eisenstein series $E_2$, $E_4$ and $E_6$ (see Appendix \ref{secn:appa}
for our conventions and definitions). This is
indeed what happens. In fact we have
 \begin{align}
h_0&=\frac{1}{4}\big(4m^2-s^2\big)~,\phantom{\Big|}\label{h02*}\\
h_1&=\frac{1}{2^6\cdot3}\big(4m^2-s^2\big)\big(4m^2+4p-s^2\big)\,E_2~,\phantom{\Big|}\label{h12*}\\
h_2&=\frac{1}{2^9\cdot3^2\cdot5}\big(4m^2-s^2\big)\big(4m^2+4p-s^2\big)
\phantom{\Big|}
\nonumber\\&\hspace{0.7cm}\times~\Big[5\big(4m^2+6p-s^2\big)E_2^2
+\big(4m^2+6p-13s^2\big)E_4\Big]~,\phantom{\Big|}\label{h22*}\\
h_3&=\frac{1}{2^{14}\cdot3^3\cdot5\cdot7}\big(4m^2-s^2\big)\big(4m^2+4p-s^2\big)
\phantom{\Big|}
\nonumber\\&\hspace{0.7cm}\times~\Big[
35\big(80m^4+272m^2p+240p^2-40m^2s^2-68ps^2+5s^4)E_2^3\phantom{\Big|}
\nonumber\\&\hspace{1.5cm}
+84\big(16m^4+64m^2p+60p^2-56m^2s^2-136ps^2+13s^4\big)E_4\,E_2\phantom{\Big|}
\nonumber\\&\hspace{1.5cm}
+\big(176m^4+944m^2p+960p^2-1816m^2s^2-4556ps^2+3323s^4\big)E_6\Big]~.\label{h32*}
 \end{align}
By using the small $q$ expansion of the Eisenstein series one can check that the explicit
instanton contributions we have computed using localization techniques are correctly recovered
from the previous formulas. 
We stress that the fact that the various instanton terms nicely combine into expressions
involving only the Eisenstein series is not obvious a priori and is a very strong 
{\emph{a posteriori}} test on the numerical
coefficients appearing in (\ref{h0q})-(\ref{h3q}). It is quite remarkable that the explicit instanton results at low $k$ can be nicely extrapolated 
and allow to reconstruct modular forms from which, by expanding in powers of $q$, one can obtain 
the contributions at any instanton number.

We can also organize the generalized prepotential according to (\ref{Fng}) and
obtain the amplitudes $F^{(n,g)}$ as a series in inverse powers
of $a^2$ with coefficients that are polynomials in $E_2$, $E_4$ and $E_6$. 
The first few of such amplitudes are:
 \begin{align}
  F^{(0,0)}&= m^2\log\frac{2a}{\Lambda}-\frac{m^4\,E_2}{48\,a^2}-
\frac{m^6(5E_2^2+E_4)}{5760\,a^4}
-\frac{m^8(175E_2^3+84E_2E_4+11E_6)}{2903040\,a^6}+ \cdots~,\label{F00}\\
F^{(1,0)}&=-\frac{1}{4}\log\frac{2a}{\Lambda}
+\frac{m^2\,E_2}{96\,a^2}+\frac{m^4(E_2^2+E_4)}{1536\,a^4}+
\frac{m^6(175E_2^3+336E_2E_4+8E_6)}{2903040\,a^6}+\cdots\label{F01}\\
F^{(0,1)}&=-\frac{m^2\,E_2}{48\,a^2}-\frac{m^4(5E_2^2+E_4)}{2304\,a^4}
-\frac{m^6(11E_2^3+6E_2E_4+E_6)}{41472\,a^6}+\cdots\phantom{\Bigg|}\label{F10}
\\
F^{(2,0)}&=-\frac{E_2}{768\,a^2}-\frac{m^2(5E_2^2+9E_4)}{30720\,a^4}+
\frac{m^4(175E_2^3+588E_2E_4+559E_6)}{7741440\,a^6}+\cdots\phantom{\Bigg|}\label{F20}\\
F^{(1,1)}&=\frac{E_2}{192\,a^2}+\frac{m^2(25E_2^2+17E_4)}{23040\,a^4}+
\frac{m^4(385E_2^3+798E_2E_4+213E_6)}{1935360\,a^6}+\cdots\phantom{\Bigg|}\label{F11}\\
F^{(0,2)}&=-\frac{m^2(5E_2^2+E_4)}{3840\,a^4}-
 \frac{m^4(160E_2^3+93E_2E_4+17E_6)}{414720\,a^6}+\cdots\label{F02}
 \end{align}
Terms with higher values of $n$ and $g$ can be systematically generated without any 
difficulty from the $h_\ell$'s given in (\ref{h02*})-(\ref{h32*}). 
The first term $F^{(0,0)}$ represents the prepotential of the $\cN=2^*$ SU(2) gauge theory and 
its expression (\ref{F00}) agrees with that found in \cite{Minahan:1997if} from the 
SW curve (see also \cite{Billo':2011pr}). 
The other terms are generalizations of those considered
in \cite{Huang:2011qx} and more recently in \cite{KashaniPoor:2012wb} in particular limits
($m\to0$ or $\epsilon_2\to 0$) where they drastically simplify.

\subsection{Recursion relations}
\label{subsecn:recn2*}
The generalized prepotential (\ref{fexp}) is clearly holomorphic by construction but does not have nice transformation properties under the modular group since the coefficients 
$h_\ell$ explicitly depend on the second Eisenstein series $E_2$ which is not a good modular function. Indeed, under
\begin{equation}
 \tau_0\to\frac{a\tau_0+b}{c\tau_0+d}\quad \mbox{with}~~ad-bc=1~,
\label{duality}
\end{equation}
$E_2$ transforms inhomogeneously as follows
\begin{equation}
 E_2(\tau_0)\to (c\tau_0+d)^2 E_2(\tau_0)+ (c\tau_0+d)\,\frac{6c}{\pi\ii}~.
\label{E2transf}
\end{equation}
Therefore, in order to have good modular properties we should replace everywhere 
$E_2$ with the shifted Eisenstein series $\widehat E_2= E_2+\frac{6}{\pi\ii(\tau_0-\bar\tau_0)}$ 
at the price, however, of loosing holomorphicity. 
This fact leads to the so-called holomorphic anomaly equation
\cite{Bershadsky:1993ta,Bershadsky:1993cx} (see also \cite{Huang:2010kf}). 
On the other hand, in the limit $\bar \tau_0 \to\infty$, holding $\tau$ fixed so that 
$\widehat E_2 \to E_2$, we recover holomorphicity but loose good modular properties and obtain the so-called modular anomaly equation \cite{Minahan:1997ct}. This equation can be rephrased
in terms of a recursion relation satisfied by the $h_\ell$ coefficients which allows to completely
fix their dependence on $E_2$. 

To this aim let us consider the explicit expressions (\ref{h02*})-(\ref{h32*})
and compute the derivatives of $h_\ell$ with respect to $E_2$. With simple algebra we find
\begin{equation}
 \frac{\partial h_\ell}{\partial E_2}=\frac{\ell}{12}\,\sum_{i=0}^{\ell-1}h_i h_{\ell-i-1}
+\frac{\ell(2\ell-1)}{12}\,\epsilon_1\epsilon_2\,h_{\ell-1}
\label{recn2*}
\end{equation}
with the initial condition
\begin{equation}
 \frac{\partial h_0}{\partial E_2}=0~.
\label{dh0}
\end{equation}
We have explicitly checked this relation for several values of $\ell>3$; 
we can thus regard it as a distinctive property of the $\epsilon$-deformed $\cN=2^*$ low-energy
theory.
Notice that in the limit $\epsilon_\ell\to 0$, (\ref{recn2*}) reduces to
the recursion relation satisfied by the coefficients of the prepotential of the 
$\cN=2^*$ SU(2) theory found in \cite{Minahan:1997if,Billo':2011pr} from the SW curve, 
and that the linear term in the right hand side disappears in the so-called Nekrasov-Shatashvili 
limit \cite{Nekrasov:2009rc} where one of the two deformation parameters vanishes and a
generalized SW curve can be introduced \cite{Poghossian:2010pn,Fucito:2011pn}.

Eq.~(\ref{recn2*}) can be formulated also as a recursion relation for the amplitudes 
$F^{(n,g)}$ defined in (\ref{Fng}). To see this, let us first extract the $a$-dependence 
and, in analogy with (\ref{fexp}), write
\begin{equation}
 F^{(n,g)}= -\sum_{\ell=1}^\infty
\frac{f^{(n,g)}_\ell}{2^{\ell+1}\ell}\frac{1}{a^{2\ell}}~,
\end{equation}
so that
\begin{equation}
 h_\ell=\sum_{n,g} f^{(n,g)}_\ell(\epsilon_1+\epsilon_2)^{2n}(\epsilon_1\epsilon_2)^g~.
\label{hfng}
\end{equation}
Notice that the coefficients $f^{(n,g)}_\ell$, which are polynomials in the hypermultiplet mass
$m$, have mass dimensions $2(1+\ell-n-g)$; therefore $f^{(n,g)}_\ell=0$ if $n+g>\ell+1$
(this condition can be easily checked on the explicit expressions (\ref{F00})-(\ref{F20}). 
These definitions must be supplemented by the ``initial conditions''
\begin{equation}
f^{(0,0)}_0=m^2~,~~f^{(1,0)}_0=-\frac{1}{4}~,~~f^{(0,1)}_0=0
\label{inti}
\end{equation}
which are obtained from (\ref{h02*}).
Inserting (\ref{hfng}) in the recursion relation (\ref{recn2*}), we obtain
\begin{equation}
 \frac{\partial f^{(n,g)}_\ell}{\partial E_2}= \frac{\ell}{12}\sum_{n1,n2;g_1,g_2}
{}^{\!\!\!\!\!\!\!\!\!\!\prime}~~\Big(
\sum_{i=0}^{\ell-1}
f^{(n_1,g_1)}_if^{(n_2,g_2)}_{\ell-i-1}\Big) +\frac{\ell(2\ell-1)}{12}\,f^{(n,g-1)}_{\ell-1}
\label{recfng}
\end{equation}
where the $^\prime$ means that the sum is performed over all $n_1$, $n_2$, $g_1$ and $g_2$ such that $n_1+n_2=n$ and $g_1+g_2=g$. 
Eq.~(\ref{recfng}) shows that the coefficients $f^{(n,g)}_\ell$ and hence the
amplitudes $F^{(n,g)}$ are recursively related to those with lower values of $n$ and $g$, similarly to what can be deduced from the holomorphic anomaly equation
\cite{Huang:2011qx}.

\section{The $\cN=2$ SU(2) theory with $N_f=4$}
\label{secn:nf4}
We now consider the $\cN=2$ SU(2) theory with $N_f$ flavor hypermultiplets in the fundamental representation
of the gauge group. When $N_f=4$ the 1-loop $\beta$-function vanishes and the conformal invariance is 
broken only by the flavor masses $m_f$ ($f=1,...,4)$. Furthermore there are non-perturbative
effects due to instantons which are nicely encoded in the exact SW solution 
\cite{Seiberg:1994aj}. We now discuss the generalizations of these effects induced by the 
$\epsilon_\ell$ deformation parameters in the Nekrasov
partition function, following the same path described in the previous section for the $\cN=2^*$ theory.  
\subsection{Instanton partition functions}
Using localization techniques, we can express the instanton partition functions $Z_k$
of the $N_f=4$ theory as sums of terms related to pairs of Young tableaux. 
Referring again the Appendix A of \cite{Billo:2012st} 
for details, at $k=1$ we have the following two contributions
\begin{equation}
 \begin{aligned}
  Z_{(\Yfund,\bullet)}&=\frac{1}{(-\epsilon_1)
(-\epsilon_2) a_{12} (a_{21}-\epsilon_1-\epsilon_2)}\prod_{f=1}^4(a_1+\widetilde m_f)~,\\
Z_{(\bullet,\Yfund)}&=\frac{1}{(-\epsilon_1)
(-\epsilon_2)a_{21}(a_{12}-\epsilon_1-\epsilon_2)}\prod_{f=1}^4(a_2+\widetilde m_f)
 \end{aligned}
\label{Z1nf4}
\end{equation}
where
\begin{equation}
 \widetilde m_f = m_f+\frac{\epsilon_1+\epsilon_2}{2}
\label{tildemf}
\end{equation}
is the equivariant hypermultiplet mass in the $\epsilon$-background%
\footnote{Note that in \cite{Billo:2012st} the hypermultiplet masses were denoted $-m_f$.}.
Summing the two terms (\ref{Z1nf4}) and setting $a_1=-a_2=a$, we get
\begin{equation}
\begin{aligned}
 Z_1 &= Z_{(\Yfund,\bullet)}+Z_{(\bullet,\Yfund)}\\
&=-\frac{1}{2\eu\ed}\Bigg[a^2 + \Smumu + (\eu + \ed) \Smu + \frac{3}{4} (\eu+\ed)^2 +
\frac{4\,\Pf m}{4 a^2 - (\eu + \ed)^2}\Bigg]
\end{aligned}
\label{z1}
\end{equation}
where $\Pf m\equiv m_1m_2m_3m_4$. 
The partition function $Z_1$ contains a part 
proportional to $a^2$, a part which does not depend on $a$ and 
a part containing $a^2$ in the denominator.
In all $a$-dependent terms the flavor masses $m_f$ always occur in SO(8)-invariant combinations 
and thus in all such terms we can always express the mass dependence using the following quadratic, 
quartic and sextic SO(8) invariants:
\begin{equation}
 \label{invdef}
 \begin{aligned}
 R & = \frac 12 \sum_f m_f^2
~,\\
  T_1 & = \frac{1}{12} \sum_{f<f'} m_f^2 m_{f'}^2 - \frac{1}{24} \sum_f m_f^4
~,\\
 T_2 & = -\frac{1}{24} \sum_{f<f'} m_f^2 m_{f'}^2 + \frac{1}{48} \sum_f m_f^4 
-\frac 12 \Pf m
~,\\
N & = \frac{3}{16} \sum_{f<f'<f''} m_f^2 m_{f'}^2 m_{f''}^2 - \frac{1}{96} 
 \sum_{f\not= f'} m_f^2 m_{f'}^4 + \frac{1}{96} \sum_f m_f^6~.
\end{aligned} 
\end{equation}
The $a$-independent terms in (\ref{z1}), instead, are not invariant under the SO(8) flavor group. 
Indeed, neither $\Smu$ nor $\Smumu$ respect this invariance. 
Notice, however, that $a$-independent terms in the partition functions, and hence in the prepotential, 
are irrelevant for the gauge theory dynamics, and thus can be neglected. We will 
always do so in presenting our results.

The explicit expressions of the partition functions $Z_k$ for $k>1$ are rather cumbersome; nevertheless
it is possible to write the (grand-canonical) instanton partition function in a quite compact way
using the connection with the Young tableaux. Indeed, denoting as $x$ the instanton counting
parameter and using the same notations introduced in
Section \ref{subsecn:inst2*}, we have
\begin{equation}
\begin{aligned}
 Z_{\mathrm{inst}} &= \sum_{k=0}^\infty x^k\,Z_k
 \\&
 =\sum_{(Y_1,Y_2)} x^{|Y|}\,\prod_{i,j=1}^\infty
\Bigg[\prod_{u,v=1}^2\frac{a_{uv}+\epsilon_1(i-1)-\epsilon_2j}{a_{uv}+\epsilon_1(i-1-\widetilde k_{vj})
-\epsilon_2(j-k_{ui})}\\
&~~~~~~~~~~~~~~~~\times ~\prod_{u=1}^2\prod_{f=1}^4\frac{a_{u}+\widetilde m_f+
\epsilon_1(i-1)-\epsilon_2(j-k_{ui})}{a_{u}+\widetilde m_f+\epsilon_1(i-1)-\epsilon_2j}\Bigg]~.
\end{aligned}
\label{znonf4}
\end{equation}
Here the second line represents the contribution of the gauge vector multiplet and the last line that 
of the fundamental hypermultiplets. {From} this expression, 
by selecting the appropriate Young tableaux, one
can obtain the various terms of the instanton partition function and their dependence on the 
$\epsilon_\ell$ parameters.

\subsection{Perturbative part}
Also in the $N_f=4$ theory one can deduce the perturbative contribution to the partition 
function from the ``universal'' factor of the grand-canonical 
instanton partition function (\ref{znonf4}), namely
\begin{equation}
\prod_{i,j=1}^\infty\,\Bigg[\mathop{\prod_{u,v=1}^2}_{u\not=v}
(a_{uv}+\epsilon_1(i-1)-\epsilon_2j)
\prod_{u=1}^2\prod_{f=1}^4\frac{1}{(a_{u}+\widetilde m_f+\epsilon_1(i-1)-\epsilon_2j)}\Bigg]
\label{Z1loopnf4}
\end{equation}
which, of course, requires a suitable interpretation and regularization. Using this observation 
and following again \cite{Nekrasov:2002qd,Nekrasov:2003rj}, we can write the perturbative part of 
the prepotential as
\begin{equation}
 F_{\mathrm{pert}}=\eu\ed\sum_{i,j=1}^\infty\Bigg[\mathop{\sum_{u,v=1}^2}_{u\not=v}
\log\frac{a_{uv}+\epsilon_1(i-1)-\epsilon_2j}{\Lambda}
-\sum_{u=1}^2\sum_{f=1}^4\log\frac{a_{u}+\widetilde m_f+\epsilon_1(i-1)-\epsilon_2j}{\Lambda}\Bigg]
\end{equation}
which, after using (\ref{log}) and summing over $i$ and $j$, becomes
\begin{equation}
 F_{\mathrm{pert}}=\eu\ed\Bigg[\mathop{\sum_{u,v=1}^2}_{u\not=v}\gamma_{\eu\ed}(a_{uv}) 
-\sum_{u=1}^2\sum_{f=1}^4
\gamma_{\eu\ed}(a_{u}+\widetilde m_f)\Bigg]
\label{f1loopnf4}
\end{equation}
where the function $\gamma_{\eu\ed}$ is defined in (\ref{gammae1e2}). Expanding the latter 
for small values of $\eu$ and $\ed$ and discarding as usual $a$-independent terms, we obtain
\begin{eqnarray}
F_{\mathrm{pert}}& =& -\log 16\ a^2 + \frac{1}{2}\big(4 R -s^2 + p\big)
 \log\frac{a}{\Lambda}\nonumber\\
&& ~~~\,- \frac{1}{96\,a^2}\Big[16 R^2 - 96 T_1 - 8 R (s^2 - 2p) + s^4 - 4 s^2 p + 3 p^2\Big]\label{preptoh}\\
&& ~~~\,- \frac{1}{3840\,a^4}\Big[128 R^3 - 1920 R T_1 + 768 N - 160 (R^2- 6 T_1)(s^2 - 2p)\nonumber\\
&&~~~~~~ + R(56 s^4 - 224 s^2 p + 192 p^2)
- 6 s^6 + 36 s^4p - 63 s^2 p^2 + 30 p^3\Big] + \cO\left(a^{-6}\right)\phantom{\Big|}\nonumber
\end{eqnarray}
where $s$ and $p$ are as in (\ref{sp}). One can easily check that in the limit $\epsilon_\ell\to0$ this expression correctly reproduces the 1-loop prepotential of the $\cN=2$ SU(2) $N_f=4$ theory (see for instance \cite{D'Hoker:1999ft}). 
Notice the presence of a term proportional to $a^2$ which corresponds to a finite 1-loop renormalization of the gauge coupling constant.

\subsection{Generalized prepotential}
We now consider the instanton corrections to the prepotential of the $N_f=4$ theory which is given by
\begin{equation}
 F_{\mathrm{inst}}= -\eu\ed \log Z_{\mathrm{inst}} = \sum_{k=1}^\infty
x^k\,F_k~.
\end{equation}
At $k=1$, up to $a$-independent terms, from (\ref{Z1nf4}) we simply get
\begin{equation}
F_1= -\eu\ed ~Z_1= \frac{a^2}{2}-\frac{2(T_1+2T_2)}{4a^2-s^2}
\label{f1nf4}
\end{equation}
The presence of an $a^2$-term in $F_1$ signals that there is a renormalization of the
gauge coupling constant at the 1-instanton level. Actually, the same thing occurs also at higher 
instanton numbers; indeed we find
\begin{equation}
F_2= \frac{13\,a^2}{64}+\cdots~,~~F_3=\frac{23\,a^2}{192}+\cdots~,~~
F_4=\frac{2701\,a^2}{32678}+\cdots
\end{equation}
Combining these contributions with the perturbative one given by the first term in
(\ref{f1loopnf4}) and adding also the classical tree-level prepotential $F_{\mathrm{cl}}=\log x
\,a^2$, we obtain
\begin{equation}
 \Big(\log x -\log 16+
\frac{1}{2}\,x+\frac{13}{64}\,x^2+\frac{23}{192}\,x^3+\frac{2701}{32768}\,x^4+\cdots\Big)a^2
\equiv \log q\, a^2
\label{qx}
\end{equation}
that is a redefinition of the instanton expansion parameter and hence of the gauge coupling constant.
By inverting this relation we get \cite{Grimm:2007tm}
\begin{equation}
 x=16\, q\,\big(1 - 8 q+44 q^2-192 q^3+\cdots) = \frac{\theta_2(q)^4}{\theta_3(q)^4}
\label{xqtheta}
\end{equation}
(see Appendix \ref{secn:appa} for some properties of the Jacobi $\theta$-functions). 
As pointed out for instance in \cite{Billo:2010mg}, by setting 
\begin{equation}
 q=\ee^{\ii\pi\tau_0}	\qquad\mbox{with}\qquad \tau_0=\frac{\theta}{\pi}+\ii\,\frac{8\pi}{g^2}~,
\label{qnf4}
\end{equation}
one can show that $\tau_0$ is precisely the modular parameter of the SW curve 
for the SU(2) $N_f=4$ theory appearing in the original paper \cite{Seiberg:1994aj}. 
This gauge coupling constant receives only corrections proportional to the hypermultiplet masses 
and as such is the strict analogue%
\footnote{Except for a customary overall factor of 2.} of the gauge coupling constant 
$\tau_0$ of the $\cN=2^*$ theory considered in Section~\ref{secn:N2star}. On the other hand, using (\ref{xqtheta}) one can show that $x$ is a cross-ratio of the four roots of the original SW curve \cite{Seiberg:1994aj}. 
{From} now on we always present the results for the $N_f=4$ theory in terms of $q$.

Expanding the complete prepotential $F_{\mathrm{pert}}+F_{\mathrm{inst}}$ as in (\ref{fexp}),
the first coefficients $h_\ell$ up to three instantons turn out to be
\begin{align}
h_0 = &~\frac{1}{2}\big(4R - s^2 +p\big)~,\label{h0nf4}\\
h_1 = &~\frac{1}{24}\Big[16(R^2 - 6 T_1) - 8 R (s^2 - 2 p) + s^4 - 4 s^2 p + 3 p^2\Big] 
+ 32(T_1 + 2 T_2)\,q\notag\\
&~- \Big[16 (R^2 + 6 T_1) - 8 R (s^2 - 2 p) + s^4 - 4 s^2 p + 3 p^2\Big]\,q^2\notag\\
&~+ 128\big(T_1 + 2 T_2\big)\,q^3 + \cO(q^4)~,\label{h1nf4}\\
h_2 =&~ \frac{1}{240}\Big[128 R^3 - 1920 R T_1 + 768 N - 160 (R^2 - 6 T_1) (s^2-2p) \notag\\
&~~~~~~~ + R(56s^4 - 224 s^2 p + 192 p^2) - 6 s^6 + 36 s^4 p - 63 s^2 p^2 + 30 p^3\Big]\notag\\
&~+ 32 s^2 (T_1 + 2 T_2)\, q \notag\\
&~+ \Big[768 N + 384 R T_1 - 64 (R^2 + 6 T_1)(s^2 - p) + 8R (4 s^4 - 12 s^2 p + 5 p^2)\notag\\
&~~~~~~~ - 4 s^6 + 20 s^4 p - 25 s^2 p^2 + 6 p^3\Big]\, q^2\notag\\
&~- 128(T_1 + 2 T_2)\big(16R - 11 s^2 + 16 p\big)\,q^3 + \cO(q^4)~.\label{h2nf4}
\end{align}
We worked out also the expressions for a few other $h_\ell$'s with $\ell>3$ which however
become rapidly very cumbersome and thus we do not write them here 
(see for example Appendix \ref{secn:appb} where we give the detailed form of $h_3$). 

As in the $\cN=2^*$ model, also in the $N_f=4$ theory we can view the previous 
results as the first instanton
terms of the expansion of (quasi) modular forms in powers of $q$. In this case, however,
besides the Eisenstein series $E_2$, $E_4$ and $E_6$, also the Jacobi $\theta$-functions
$\theta_2$ and $\theta_4$ are needed. Matching the $q$-expansion of these modular functions
with the explicit results (\ref{h0nf4})-(\ref{h2nf4}), guided by what we already 
obtained in the $\epsilon_\ell\to 0$ limit \cite{Billo':2011pr} which suggests the existence
of a recursion relation that fixes the $E_2$ dependence, we find%
\footnote{As compared to the results presented in \cite{Billo':2011pr,Billo:2010mg}, here all
masses have been rescaled by a factor of $\sqrt{2}$, that is $m_f^{\mathrm{there}}= \sqrt{2}\,m_f^{\mathrm{here}}$ for all $f$.} 
\begin{align}
h_0 = &~\frac{1}{2}\big(4R - s^2 +p\big)~,\label{h0nf4a}\\
 h_1=&~
\frac{1}{24}\big(4R -s^2 +p\big)\big(4R -s^2 +3p\big)\, E_2-4 \big(T_1 \,\theta_4^4- T_2\,\theta_2^4\big)~,\label{h1nf4a}\\
h_2=&~\frac{1}{144}\big(4R - s^2 +p\big)\big(4R - s^2 +3p\big)\big(4R - s^2 +4p\big)\,E_2^2
\notag\\&~
-\frac{4}{3}\big(4R -s^2 +4p\big)\big(T_1 \,\theta_4^4-T_2 \,\theta_2^4\big)\,E_2\notag\\
&~+\frac{1}{720}\Big[64 R^3-80 R^2(3s^2-4p)+4R(27s^4-88s^2p+49p^2)\notag\\
&~~~~~~~~~~~~-13s^6+68s^4p-94s^2p^2+30p^3+2304 N\Big]\,E_4\notag\\
&~-\frac{8}{3}\big(R -s^2 +p\big)\Big[T_1 \,\theta_4^4\big(2\theta_2^4+\theta_4^4\big)
+T_2 \,\theta_2^4\big(\theta_2^4+2\theta_4^4\big)\Big]~.
\label{h2nf4a}
\end{align}
We have checked that a similar pattern occurs also in other $h_\ell$'s with $\ell>3$
(see Appendix \ref{secn:appb}). 

Organizing the complete prepotential as an expansion in powers of $s^2$ and $p$ as in (\ref{Fng}),
we can obtain the amplitudes $F^{(n,g)}$, the first few of which are
\begin{align}
 F^{(0,0)}=&~ 2 R \log\frac{a}{\Lambda}-\frac{R^2E_2}{6a^2}+\frac{T_1\theta_4^4-T_2\theta_2^4}{a^2}
-\frac{R^3(5E_2^2+E_4)}{180\,a^4}-\frac{NE_4}{5\,a^4}\notag\\
&~+\frac{RT_1\theta_4^4(2E_2+2\theta_2^4+\theta_4^4)}{6\,a^4}-
\frac{RT_2\theta_2^4(2E_2+2\theta_4^4+\theta_2^4)}{6\,a^4}
+\cdots~,\label{f00nf4}\\
F^{(1,0)}=&~-\frac{1}{2}\log \frac{a}{\Lambda}+\frac{R E_2}{12\,a^2}
+\frac{R^2(E_2^2+E_4)}{48\,a^4}\notag\\
&~-\frac{T_1\theta_4^4(E_2+4\theta_2^4+2\theta_4^4)}{12\,a^4}+
\frac{T_2\theta_2^4(E_2-4\theta_4^4-2\theta_2^4)}{12\,a^4}+\cdots~,\label{f10nf4}\\
F^{(0,1)}=&~\frac{1}{2}\log\frac{a}{\Lambda}-\frac{RE_2}{6\,a^2}
-\frac{R^2(2E_2^2+E_4)}{36\,a^4}\notag\\
&~+\frac{T_1\theta_4^4(2E_2+2\theta_2^4+\theta_4^4)}{6\,a^4}-
\frac{T_2\theta_2^4(2E_2-2\theta_4^4-\theta_2^4)}{6\,a^4}+\cdots~,\label{f01nf4}\\
F^{(2,0)}=&~-\frac{E_2}{96a^2}-\frac{R(5E_2^2+9E_4)}{960\,a^4}+\cdots~,\label{f20nf4}\\
F^{(1,1)}=&~\frac{E_2}{24a^2}+\frac{R(10E_2^2+11E_4)}{360\,a^4}+\cdots~,\label{f11nf4}\\
F^{(0,2)}=&~-\frac{E_2}{32a^2}-\frac{R(95E_2^2+49E_4)}{2880\,a^4}+\cdots~,\label{f02nf4}\\
F^{(3,0)}=&~\frac{5E_2^2+13E_4}{11520\,a^4}\cdots~,\qquad\qquad~ F^{(2,1)}=-\frac{10E_2^2+17E_4}{2880\,a^4}\cdots~,\label{f30nf4}\\
F^{(1,2)}=&~\frac{95E_2^2+94E_4}{11520\,a^4}\cdots~,\qquad\qquad F^{(0,3)}=-\frac{2E_2^2+E_4}{384\,a^4}\cdots~.\label{f03nf4}
\end{align}
One can easily check that $F^{(0,0)}$ in (\ref{f00nf4}) completely agrees with the prepotential of the
$\cN=2$ SU(2) $N_f=4$ gauge theory as derived for example in \cite{Billo':2011pr};
notice also that in the massless limit, {\it i.e.} 
$R,N,T_1,T_2\to 0$, these amplitude drastically simplify and
precisely match the results presented in \cite{Huang:2011qx}.
It is also interesting to observe that in the Nekrasov-Shatshvili limit where one of the $\epsilon_\ell$'s vanishes, the amplitudes $F^{(n,0)}$ of the $N_f=4$ theory reduce to those
of the $\cN=2^*$ theory upon setting $T_1=T_2=N=0$ and $R=m^2$. Indeed,
with these positions and rescaling $a\to 2a$, the amplitudes $F^{(n,0)}$ in 
(\ref{f00nf4})-(\ref{f03nf4}) become twice the corresponding amplitudes of the $\cN=2^*$ theory
given in (\ref{F00})-(\ref{F02}). This simple relation does not hold away from the Nekrasov-Shatashvili
limit: the amplitudes $F^{(n,g)}$ with $g\not=0$ are in fact intrinsically different in the
two theories, as a consequence of their different gravitational structure. 

\subsection{Recursion relations}
Looking at the explicit expressions (\ref{h0nf4a})\,-\,(\ref{h2nf4a}), it is not difficult to realize
that the $E_2$ dependence of the $h_\ell$'s is quite simple and exhibits a recursive pattern.
This points again to the existence of a recursion relation among the $h_\ell$'s involving
their derivatives with respect to $E_2$. Indeed, following the same steps described in Section~\ref{subsecn:recn2*}, one can check that
\begin{equation}
 \frac{\partial h_\ell}{\partial E_2}=\frac{\ell}{6}\,\sum_{i=0}^{\ell-1}h_i h_{\ell-i-1}
+\frac{\ell(2\ell-1)}{6}\,\epsilon_1\epsilon_2\,h_{\ell-1}
\label{recnf4}
\end{equation}
with the initial condition
\begin{equation}
 \frac{\partial h_0}{\partial E_2}=0~.
\label{dh0nf4}
\end{equation}
This recursion relation has exactly the same structure of that of the $\cN=2^*$ theory given
in (\ref{recn2*}), the only difference being in the numerical coefficients. Notice that 
the coefficients of the quadratic terms can be matched by a 
rescaling of the $h_\ell$'s, but those of the linear terms proportional to $\eu\ed$
remain different for the two theories; this is another signal of their intrinsically different
behavior when a generic $\epsilon$-background is considered.

Finally, if we expand the $h_\ell$'s as in (\ref{hfng}) we can reformulate the
recursion relation (\ref{recnf4}) in terms of the partial amplitudes $f^{(n,g)}_\ell$ and get
\begin{equation}
 \frac{\partial f^{(n,g)}_\ell}{\partial E_2}= \frac{\ell}{6}\sum_{n_1,n_2;g_1,g_2}
{}^{\!\!\!\!\!\!\!\!\!\!\prime}~~\Big(
\sum_{i=0}^{\ell-1}
f^{(n_1,g_1)}_if^{(n_2,g_2)}_{\ell-i-1}\Big) +\frac{\ell(2\ell-1)}{6}\,f^{(n,g-1)}_{\ell-1}
\label{recfngnf4}
\end{equation}
where the $^\prime$ means that the sum is performed over all $n_1$, $n_2$, $g_1$ and $g_2$ such that $n_1+n_2=n$ and $g_1+g_2=g$. This relation has to be supplemented by the initial conditions
\begin{equation}
f^{(0,0)}_0=2R~,~~f^{(1,0)}_0=-\frac{1}{2}~,~~f^{(0,1)}_0=\frac{1}{2}
\label{intinf4}
\end{equation}
obtained from (\ref{h0nf4a}). 

In the next section we will analyze the implications of the recursion relations and in particular 
their consequences on the modular transformation properties of the generalized prepotential.

\section{Modular anomaly equations and S-duality}
\label{secn:sdual}
The fact that the generalized prepotential can be written in terms
of (quasi) modular functions of the bare coupling constant allows to explore the modularity
properties of the deformed theory and study how the $\mathrm{Sl}(2,\mathbb Z)$ symmetry
of the microscopic high-energy theory is realized in the effective low-energy theory.
In the following we will concentrate on the SU(2) $N_f=4$ theory,
even if our analysis and our methods can be equally well applied to the $\cN=2^*$ SU(2) model.
Furthermore, we will focus on the generator $\cS$ of the modular group, corresponding to the
following transformation of the bare coupling constant
\begin{equation}
 \cS\,:~~\tau_0~\to~-\frac{1}{\tau_0}~.
\label{Stau0}
\end{equation}
Notice that this action implies that the instanton counting parameter $x$ in the Nekrasov's partition
of the $N_f=4$ theory, defined in (\ref{xqtheta}), transforms as follows
\begin{equation}
 \cS\,:~~x~\to~1-x~,
\label{Sx}
\end{equation}
as one can readily check from the properties of the Jacobi $\theta$-functions. This type of transformation is consistent with the interpretation of $x$ as a cross-ratio, a fact that is
also exploited in the AGT correspondence with the Liouville conformal blocks \cite{Alday:2009aq}.
As discussed in \cite{Seiberg:1994aj}, in the $N_f=4$ theory the modular group acts
with triality transformations on the mass invariants (\ref{invdef}); in particular one has
\begin{equation}
\cS\,:~~R~\to~R\quad,\quad T_1~\to~T_2\quad,\quad T_2~\to~T_1\quad,\quad N~\to~N~.
\label{ST}
\end{equation}

In the deformed theory, these rules have 
to be supplemented by those that specify how $\mathrm{Sl}(2,\mathbb Z)$ acts on the 
equivariant deformation parameters.
Adapting to the present case the considerations made in \cite{Dimofte:2011jd}
for the $\epsilon$-deformed conformal Chern-Simons theory in three dimensions, we assume that
$\cS$ simply exchanges $\eu$ and $\ed$ with each other%
\footnote{Note that this rule is consistent with the interpretation of $\eu$ and $\ed$
as fluxes of (complex) combinations of NS-NS and R-R 3-form field strengths in a Type IIB 
string theory realization which rotate among themselves under S-duality.}. 
In particular this means that $s=\eu+\ed$ and $p=\eu\ed$ are left unchanged, {\it i.e.}
\begin{equation}
\cS\,:~~s~\to~s\quad,\quad p~\to~p~.
\label{Ssp}
\end{equation}
Using the rules (\ref{Stau0})\,-\,(\ref{Ssp}) and the modular properties of the Eisenstein series and
Jacobi $\theta$-functions, it is easy to show that the
coefficients $h_{\ell}$ of the generalized prepotential 
(see (\ref{h2nf4a}) and Appendix~\ref{secn:appb}) 
transform as quasi-modular forms of weight $2\ell$ with anomalous terms due to the presence of 
the second Eisenstein series $E_2$, namely
\begin{equation}
\label{deltah}
\cS\,:~~h_ {\ell}(E_2)~\to ~
\tau_0^{2\ell}\,h_ {\ell}\Big(E_2+\frac{6}{\pi\ii\tau_0}\Big)
=\tau_0^{2\ell}\,\sum_{k=0}^{\ell}\frac{1}{k!}
\left(\frac{1}{2\pi \ii \tau_0}\right)^k D^k h_\ell(E_2)
\end{equation} 
where we have introduced the convenient notation $D\equiv 12 \partial_{E_2}$.
These transformation rules are formally identical to those of the underformed $N_f=4$ theory derived
in \cite{Billo':2011pr} from the SW curve; however, since the $h_\ell$'s satisfy a modified
recursion relation with a new term proportional to $\eu\ed$, the practical effects of 
(\ref{deltah}) in the deformed theory are different from those of the undeformed
case, as we shall see momentarily. 

To proceed let us consider the pair made by $a$ and its S-dual image $\cS(a)\equiv a_D$, on which $\cS$ acts as follows \cite{Seiberg:1994rs,Seiberg:1994aj}:
\begin{equation}
 \label{aadual}
\cS\,:~~\begin{pmatrix} a \cr
a_D \end{pmatrix}~\to~
\begin{pmatrix} 0 & ~1~ \cr
-1& ~0~\end{pmatrix}\,\begin{pmatrix} a \cr
a_D\end{pmatrix}=\begin{pmatrix}a_D \cr
-a\end{pmatrix}~.
\end{equation}
We therefore have 
\begin{equation}
 \cS^2(a)=-a~.
\label{S2a}
\end{equation}
In the SW theory this relation is enforced by taking 
\begin{equation}
 \cS(a)=\frac{1}{2\pi\ii}\,\frac{\partial\cF}{\partial a}
\label{adual}
\end{equation}
where $\cF$ is the undeformed effective prepotential which is related to its S-dual by a 
Legendre transform:
\begin{equation}
 \cF-\cS(\cF)= 2\pi\ii\,a\,\cS(a)~.
\label{legendre}
\end{equation}
It seems natural to try the same thing also in the deformed theory. 
Recalling that the effective generalized prepotential is 
\begin{equation}
 F= \pi\ii\tau_0\, a^2 
+h_0 \log\frac{a}{\Lambda}-\sum_{\ell=1}^\infty \frac{h_\ell}{2^{\ell+1}\,\ell}\,\frac{1}{a^{2\ell}}
\label{Fnf4}
\end{equation}
where the first term is the classical part, we therefore posit 
\begin{equation}
\label{ad}
\cS (a)=\frac{1}{2\pi\ii}\,\frac{\partial F}{\partial a}
= \tau_0\, a +\, \frac{1}{2\pi\ii}
\sum_{\ell=0}^\infty \frac{h_\ell}{2^{\ell}}\,\frac{1}{a^{2\ell+1}}~.
\end{equation}
Applying the S-duality rules (\ref{Stau0}) and (\ref{deltah}), we obtain
\begin{eqnarray}
\cS^2(a) &=& -\frac{{\cS}(a)}{\tau_0}
+\, \frac{1}{2\pi\ii}
\sum_{\ell=0}^\infty \frac{\tau_0^{2l}\left( h_\ell+\frac{1}{2\pi \ii \tau_0}
D h_{\ell}+{\cO}(\tau_0^{-2})\right) }{2^{\ell} \,({\cS}(a))^{2\ell+1}} 
\label{S2a0}\\
&=&-a+\frac{1}{(2\pi\ii\tau_0)^2}
\Bigg[\sum_{\ell=0}^\infty\frac{D h_{\ell}}{2^\ell}\,\frac{1}{a^{2\ell+1}}
-\sum_{\ell,n=0}^\infty\frac{(2\ell+1)\,h_\ell\,h_n}{2^{\ell+n}}\,\frac{1}{a^{2\ell+2n+3}}
\Bigg] +{\cO}(\tau_0^{-3})\nonumber~.
\end{eqnarray}
The expression in square brackets can be simplified using the recursion relation 
(\ref{recnf4}); like in the undeformed theory, the quadratic terms in
the $h$'s exactly cancel but, due to the new $\epsilon$-dependent term in the deformed 
recursion relation, an uncanceled part proportional to $\eu\ed$ remains.
This simple calculation shows that in order to enforce the relation (\ref{S2a}) when $\eu\ed\not=0$,
the standard definition (\ref{ad}) has to be modified by 
adding terms proportional to $\eu\ed$ in the right hand side.
In the sequel we will work out explicitly the first corrections 
and show how the relation (\ref{S2a}) constrains the form of $\cS(a)$. 

\subsection{S-duality at first order in $\eu\ed$}
\label{subsecn:first}
To organize the calculation, we introduce a set of generating functions 
$\varphi_\ell$ for the coefficients $h_{\ell}$ according to
\begin{equation}
\begin{aligned}
 \varphi_0 = -h_0 \log\frac{a}{\Lambda}+ \sum_{\ell=1}^\infty
\frac{ h_\ell}{2^{\ell+1}\,\ell}\frac{1}{a^{2\ell}}~,\qquad
\varphi_{\ell+1}=-\partial_a\varphi_\ell\quad\mbox{for}~~~\ell\geq 0~.
\end{aligned}
\label{phil}
\end{equation}
In particular we have the following relations with
the generalized prepotential $F$:
\begin{subequations}
 \begin{align}
  \varphi_0 &= \pi\ii\tau_0\,a^2-F~,
\label{phi0F}\\
\varphi_1 &= -2\pi\ii\tau_0\,a+\partial_a F~,
\label{phi1F}\\
\varphi_2 &= 2\pi\ii\tau_0-\partial_a^2 F \,\equiv\, 2\pi\ii\tau_0-2\pi\ii\tau~,
\label{phi2F}
 \end{align}
\end{subequations}
where in the last line we have introduced the effective coupling $\tau$.

As shown in Appendix~\ref{secn:appc}, the $\varphi_\ell$'s form a ring under the action 
of $D$ due to the generalized modular anomaly equations (\ref{recnf4}). 
This ring structure will enable us to formally express everything 
as functions of the $\varphi_\ell$'s with coefficients that may depend on the product 
$\eu\ed$. Such a dependence is a consequence of the $\eu\ed$-term in the recursion relation (\ref{recnf4}) and is the only explicit dependence on the deformation parameters that will be
relevant for our purposes, all the rest being implicit inside the $\varphi_\ell$'s 
and the $h_\ell$'s therein. 

As argued above, the definition (\ref{ad}) for $\cS(a)$, namely $\cS(a) = \tau_0 a + \varphi_1/(2\pi\ii)$,  
has to be replaced by a new one containing  terms proportional to $\eu\ed$, {\it i.e.} 
\begin{equation}
\label{Sae}
\cS (a)= \tau_0\, a +\, \frac{\varphi_1}{2\pi\ii}+\frac{X}{2\pi\ii}
\end{equation}
with
\begin{equation}
 X=\eu\ed X_1+(\eu\ed)^2X_2+\cdots
\label{X}
\end{equation}
where the $X_\ell$'s have to be determined by imposing the constraint (\ref{S2a}). 
Applying S-duality to (\ref{Sae}), it is straightforward to obtain
\begin{equation}
\label{S2ae}
\cS^2 (a)=-a+\frac{1}{2\pi\ii\tau_0}\Big[\big(\tau_0\,\cS(\varphi_1)-\varphi_1\big)
+\big(\tau_0\,\cS(X)-X\big)\Big]~;
\end{equation}
therefore, in order to satisfy the relation (\ref{S2a}), the expression in the square brackets above
must vanish. Expanding this condition in $\eu\ed$, we obtain the 
following constraints
\begin{subequations}
 \begin{align}
&\big(\tau_0\,\cS(\varphi_1)-\varphi_1\big)\Big|_0=0~,
\label{Sphi10}\\
&\big(\tau_0\,\cS(\varphi_1)-\varphi_1\big)\Big|_n+\sum_{k=1}^n
\big(\tau_0\,\cS(X_k)-X_k\big)\Big|_{n-k}=0
\label{Sphi11} 
\end{align}
\label{conditions}
\end{subequations}
\!\!where the symbol $\big|_n$ means taking the coefficient of $(\eu\ed)^n$ (notice that
the $\varphi_\ell$'s defined in (\ref{phil}) do not have any explicit dependence on $\eu\ed$ and thus
$\varphi_\ell\,\big|_n=\delta_{n0}\,\varphi_\ell$; for the same reason we also have 
$X_\ell\,\big|_n=\delta_{n0}\,X_\ell$).

Let us now compute $\cS(\varphi_1)$. Using the S-duality rules (\ref{Stau0}) and (\ref{deltah}), and
exhibiting temporarily the dependencies on $E_2$ and $a$ which are the only relevant 
ones for our purposes, we obtain
\begin{equation}
\begin{aligned}
 \tau_0\,\cS\big(\varphi_1(E_2;a)\big)&= \sum_{\ell=0}^\infty\frac{
\tau_0^{2\ell+1}h_ {\ell}\Big(E_2+\frac{6}{\pi\ii\tau_0}\Big)}{2^\ell
\left(\cS(a)\right)^{2\ell+1}}=
\sum_{\ell=0}^\infty\frac{h_ {\ell}\Big(E_2+\frac{6}{\pi\ii\tau_0}\Big)}{2^\ell
\left(a+\frac{\varphi_1}{2\pi\ii\tau_0}+\frac{X}{2\pi\ii}\right)^{2\ell+1}}\\
&=\varphi_1\Big(E_2+\frac{6}{\pi\ii\tau_0};a+\frac{\varphi_1}{2\pi\ii\tau_0}+\frac{X}{2\pi\ii}\Big)\\
&= \phantom{\Bigg|}\ee^{\frac{1}{2\pi\ii\tau_0}
(D+\zeta\,\partial_a)}\varphi_1(E_2;a)\Big|_{\zeta=\varphi_1+X}~.
\end{aligned}
\label{tSphi1}
\end{equation}
Actually, this is a particular case of the more general result
\begin{equation}
 \tau_0^\ell\,\cS(\varphi_\ell)
=\phantom{\Bigg|}\ee^{\frac{1}{2\pi\ii\tau_0}
(D+\zeta\,\partial_a)}\,\varphi_\ell\,\Big|_{\zeta=\varphi_1+X}~.
\label{tSphil}
\end{equation}
Expanding the exponential and using the relation
\begin{equation}
 (D+\zeta \partial_a)^n\varphi_\ell\Big|_{\zeta=\varphi_1+X}=
\sum_{k=0}^n(-1)^k\bin{\,n\,}{\,k\,}X^k\,(D+\zeta \partial_a)^n\varphi_{\ell+k}\Big|_{\zeta=\varphi_1}~,
\label{DX}
\end{equation}
after some straightforward algebra to rearrange the various terms,  we obtain
\begin{equation}
 \tau_0^\ell\,\cS(\varphi_\ell)= \sum_{n=0}^\infty
\frac{(-1)^n}{n!}\,\Big(\frac{X}{2\pi\ii\tau_0}\Big)^n\,\Sigma^{(\ell+n)}
\label{tSphilX}
\end{equation}
where the functions $\Sigma^{(\ell)}$ are defined by 
\begin{equation}
 \Sigma^{(\ell)}= \ee^{\frac{1}{2\pi\ii\tau_0}
(D+\zeta\,\partial_a)}\varphi_\ell\Big|_{\zeta=\varphi_1}
~.
\label{sigmas}
\end{equation}
As shown in Appendix~\ref{secn:appc}, these functions satisfy 
the remarkably simple relation
\begin{equation}
 \Sigma^{(\ell+1)} = -\frac{\tau_0}{\tau}\,\partial_a \Sigma^{(\ell)}
\label{sigmal}
\end{equation}
that is a consequence of the ring properties obeyed by the functions $\varphi_\ell$, which
in turn are due to the modular anomaly equations (\ref{recnf4}).
In view of this we can therefore rewrite (\ref{tSphilX}) as 
\begin{equation}
\tau_0^\ell\,\cS\big(\varphi_\ell\big)=\ee^{\frac{\zeta}{2\pi\ii\tau}\,\partial_a}
\,\Sigma^{(\ell)}\Big|_{\zeta=X}~.
\label{tSphifin}
\end{equation}
Note that the bare coupling $\tau_0$ initially appearing in the right hand side of (\ref{tSphil})
has been dressed into the effective coupling $\tau$. 

We now exploit this result and proceed perturbatively in $\eu\ed$ to obtain explicit expressions.
At the zeroth order in $\eu\ed$, from (\ref{tSphifin}) we have
\begin{equation}
 \tau_0\,\cS(\varphi_1)\,\Big|_0 = \Sigma^{(1)}
\,\Big|_0=\varphi_1
\label{sphi10}
\end{equation}
where the last equality follows from (\ref{sumPn1}). The
constraint (\ref{Sphi10}) is therefore identically satisfied. 
This is no surprise since we have already shown
that the relation (\ref{S2a}) must be true up to terms proportional to $\eu\ed$. At the first order
in the deformation parameters we get instead 
\begin{equation}
\begin{aligned}
 \tau_0\,\cS(\varphi_1)\,\Big|_1 &= \Sigma^{(1)}
\,\Big|_1
+\frac{X_1}{2\pi\ii\tau}\,\partial_a\Sigma^{(1)}\Big|_0
=\frac{\tau_0\,\varphi_3}{2(2\pi\ii)\tau^2}
-\frac{X_1\,\varphi_2}{2\pi\ii\tau}
\end{aligned}
\end{equation}
where we have used (\ref{sumPn1}).
Inserting this result into (\ref{Sphi11}) for $n=1$, we obtain
the following equation for $X_1$
\begin{equation}
 \cS(X_1)\,\Big|_0= \frac{X_1}{\tau}-
\frac{\varphi_3}{2(2\pi\ii)\tau^2}~.
\label{SX1}
\end{equation}
To solve it we take advantage of the S-duality relations
at the zeroth order in $\eu\ed$, which are formally the same of the SW theory.
Using (\ref{sigmal})\,-\,(\ref{sphi10}) it is easy to show that
\begin{equation}
 \cS(\varphi_2)\,\Big|_0= \frac{\varphi_2}{\tau_0\tau}\quad,\quad
 \cS(\varphi_3)\,\Big|_0= \frac{\varphi_3}{\tau^3}~,
\label{Sphi2phi3}
\end{equation}
and also that
\begin{equation}
 \cS(\tau)\,\Big|_0=-\frac{1}{\tau}~.
\label{Stau}
\end{equation}
Equipped with these results, one can check that a solution to (\ref{SX1}) is given by
\begin{equation}
 X_1= \frac{\varphi_3}{8\pi\ii\tau}~.
\label{X1fin}
\end{equation}
In conclusion%
\footnote{Note that actually the solution (\ref{X1fin}) is not unique, since in principle
one could add to $X_1$ a term $Y_1$ such that $\cS(Y_1)=Y_1/\tau$. A detailed analysis shows that
a term of this type is of the form $\alpha\,\varphi_3/(\tau^2+1)$ with $\alpha$ constant,
which has different pole structure in $\tau$ with respect to $X_1$. We therefore do
not consider this possibility and take $\alpha=0$.} we find that 
the S-dual image of $a$, which obeys the constraint (\ref{S2a}) up to terms of
order $(\eu\ed)^2$, is
\begin{equation}
\label{Saefin}
\cS (a)= \tau_0\, a +\, \frac{\varphi_1}{2\pi\ii}+\eu\ed\,\frac{\varphi_3}{4
(2\pi\ii)^2\tau}
+\cO\big((\eu\ed)^2\big)~.
\end{equation}
Observing that $\varphi_3=2\pi\ii\,\partial_a\tau$, we can rewrite the $\eu\ed$-term
above also as
\begin{equation}
 \frac{\eu\ed}{4(2\pi\ii)}\,\partial_a\log\Big(\frac{\tau}{\tau_0}\Big)~,
\label{X1fin1}
\end{equation}
so that (\ref{Saefin}) becomes
\begin{equation}
\label{Saefin1}
 2\ii\pi \,{\cS}(a) = \partial_a F  
+\frac{\eu\ed}{4}\, \partial_a 
\log\left(\frac{\tau}{\tau_0}\right)
+{\cO}\big((\eu\ed)^2\big)~.
\end{equation}
In the following we will investigate the implications of this result, while
we refer to Appendix~\ref{secn:appd} for its extension to the second order in $\eu\ed$.

\subsection{S-duality on the prepotential}

An immediate consequence of the $\epsilon$-correction in (\ref{Saefin1}) is that
$\cS(a)$ is not simply proportional to the derivative of the prepotential $F$;
thus it is natural to expect that in the deformed theory the S-dual of the prepotential
is not simply given by a Legendre transformation as it happens instead in the undeformed 
SW case (see (\ref{legendre})). 
In \cite{Galakhov:2012gw} the relation between $F$ and $\cS(F)$
has been conjectured to be a deformed Fourier transformation. Here we reach the
same conclusion, even though from a different perspective since for us all deviations 
from the undeformed theory are parametrized by the product $\eu\ed$. 

We can compute the first $\epsilon$-corrections to the relation between $F$ and $\cS(F)$
using the same methods of the previous subsection and the results given in Appendix~\ref{secn:appc}. The starting point is the relation between $\varphi_0$ and the generalized prepotential (see (\ref{phi0F}));
from this we get the useful identity
\begin{equation}
\label{FsF1}
F-\cS(F) = 2\pi\ii \,a\, \cS(a) +\frac{\pi\ii}{\tau_0}\big(\cS(a)-a\tau_0\big)^2
+\cS(\varphi_0)-\varphi_0~.
\end{equation}
Using (\ref{Sae}) and (\ref{tSphilX}) for $\ell=0$, we can easily rewrite the right hand side and
obtain
\begin{equation}
\label{FsF2}
\begin{aligned}
F-\cS(F) &= 2\pi\ii \,a\, \cS(a) +\frac{\varphi_1^2}{2(2\pi\ii)\tau_0}+\Sigma^{(0)}-\varphi_0
-\frac{X}{2\pi\ii\tau_0}\big(\Sigma^{(1)}-\varphi_1\big)\\
&~+\frac{1}{2}\,\Big(\frac{X}{2\pi\ii\tau_0}\Big)^2\big(\Sigma^{(2)}+2\pi\ii\tau_0\big)+\cdots
\end{aligned}
\end{equation}
where the dots stand for terms of order $X^3$ which are at least ${\cO}\big((\eu\ed)^3\big)$.
Notice that the difference $\Sigma^{(1)}-\varphi_1$ is of order $\eu\ed$, as one can see from
(\ref{sphi10}); thus the linear term in $X$ gives contributions ${\cO}\big((\eu\ed)^2\big)$,
like the $X^2$ term. This means that the knowledge of the first order correction $X_1$ obtained
in the previous subsection is enough to compute the correction to $F-{\cS}(F)$ at order $(\eu\ed)^2$.
This mechanism actually works at all orders, namely the $k$-th order coefficient of  
$F-{\cal S}(F)$ does not depend on $X_k$ but only on $X_j$ with $j<k$.

Using the explicit value of $X_1$ given in (\ref{X1fin}) and the expressions for the 
$\Sigma^{(\ell)}$'s given in Appendix~\ref{secn:appc} (see in particular (\ref{sumPn0})\,-\,
(\ref{sumPn2})), after straightforward algebra we obtain
\begin{equation}
\label{SFo2}
F-{\cal S}(F)
= 2\pi\ii \,a\, \cS(a) - \frac{ \epsilon_1\epsilon_2}{2}\log\Big(\frac{\tau}{\tau_0}\Big)
+(\epsilon_1\epsilon_2)^2
\Big(\frac{1}{8}\,\frac{\varphi_4}{(2\ii\pi\tau)^2}+\frac{11}{96}
\frac{\varphi_3^2}{(2\ii\pi\tau)^3}\Big)
+{\cO}\big((\epsilon_1\epsilon_2)^3\big)~.
\end{equation}
This shows that the simple Legendre transform relation (\ref{legendre}), which holds in the
underformed theory, does not work any more when $\eu\ed\not=0$, as argued,
from a different point of view, in \cite{Galakhov:2012gw}.
In the next section, however, we will show that with a suitable redefinition of the
prepotential $F$ and of the coupling constant $\tau$ it is possible to recover
the standard Legendre transform relation
also in the $\eu\ed$-deformed theory.

\section{Conclusions}
\label{secn:conl}
The explicit first-order calculation of Section~\ref{subsecn:first}
shows that $\cS(a)$ is not simply 
the derivative of the prepotential $F$. However, it is still a total derivative, as is clear from (\ref{Saefin1}).
This feature is maintained also at the second order. Indeed, as shown in Appendix~\ref{secn:appd},
the $(\eu\ed)^2$ correction $X_2$ can be chosen as
\begin{equation}
 \begin{aligned}
X_2&=\frac{1}{16}\,\frac{\varphi_5}{(2\pi\ii\tau)^2}+
\frac{23}{96}\,\frac{\varphi_3\varphi_4}{(2\pi\ii\tau)^3}+
\frac{11}{64}\,\frac{\varphi_3^3}{(2\pi\ii\tau)^4}\\
&=\partial_a\Big(\!-\frac{1}{16}\frac{\varphi_4}{(2\pi\ii\tau)^2}-
\frac{11}{192}\,\frac{\varphi_3^2}{(2\pi\ii\tau)^3}\Big)~,
 \end{aligned}
\label{X2}
\end{equation}
so that we can rewrite (\ref{Sae}) in the following way
\begin{equation}
{\cS}(a) = \frac{1}{2\pi\ii}\,\partial_a \widehat F
\label{SaFhat}
\end{equation}
with
\begin{equation}
 \widehat F = F+\frac{\eu\ed}{4}\,\log\Big(\frac{\tau}{\tau_0}\Big)
-\Big(\frac{\eu\ed}{4}\Big)^2\,\Big(\frac{\varphi_4}{(2\pi\ii\tau)^2}+
\frac{11}{12}\,\frac{\varphi_3^2}{(2\pi\ii\tau)^3}\Big)
+{\cO}\big((\eu\ed)^3\big)~.
\label{Fhat}
\end{equation}
In the deformed theory the S-duality transformation of the
effective coupling $\tau$ does not have a simple form; in fact, 
using (\ref{phi2F}) and the transformation properties of $\varphi_2$, it is easy to show that
(see (\ref{Stau2}))
\begin{equation}
 \cS(\tau) = -\frac{1}{\tau}-\eu\ed\Big(\frac{1}{2}\,\frac{\varphi_4}{(2\pi\ii)^2\tau^3}+
\frac{3}{4}\,\frac{\varphi_3^2}{(2\pi\ii)^3\tau^4}\Big)+{\cO}\big((\eu\ed)^2\big)~.
\label{Stau1}
\end{equation}
However, there is a modified effective coupling on which S-duality acts in a simple way. 
This is
\begin{equation}
 \widehat \tau \,\equiv\,\partial_a\cS(a) \,=\, \frac{1}{2\pi\ii}\,\partial_a^2\widehat F
\,=\,\tau\,\Bigg[1-\frac{\eu\ed}{4}\Big(\frac{\varphi_4}{(2\pi\ii\tau)^2}+
\frac{\varphi_3^2}{(2\pi\ii\tau)^3}\Big)+{\cO}\big((\eu\ed)^2\big)\Bigg]~.
\label{tauhat}
\end{equation}
One can easily check that
\begin{equation}
 \cS(\widehat \tau) = \partial_{\cS(a)}\cS^2(a)=-\partial_{\cS(a)}a=-\frac{1}{\widehat\tau}~.
\label{stauhat}
\end{equation}
Thus, in the effective $\epsilon$-deformed theory
it is $\widehat\tau$, and not $\tau$, that exhibits the same behavior of the bare coupling 
$\tau_0$ under S-duality. We also observe that if one uses
$\widehat\tau$, the expression of the extended prepotential
$\widehat F$ given in (\ref{Fhat}) simplifies and becomes
\begin{equation}
 \widehat F= F+\frac{\eu\ed}{4}\,\log\Big(\frac{\widehat\tau}{\tau_0}\Big)
+\Big(\frac{\eu\ed}{4}\Big)^2\,\frac{\varphi_3^2}{12(2\pi\ii\widehat\tau)^3}
+{\cO}\big((\eu\ed)^3\big)~.
\label{Fhat1}
\end{equation}
This result seems to suggest that it is possible to write the higher order corrections 
only in terms of $\varphi_3$ which, being the triple derivative of the prepotential,
is proportional to the Yukawa coupling $C_{aaa}$, the rank-three 
symmetric tensor playing a crucial r\^ole in special geometry.
Moreover, one can verify the simple Legendre transform relation
\begin{equation}
 \widehat F - \cS(\widehat F) = 2\pi\ii\,a\,\cS(a)
\label{legendre2}
\end{equation}
up to terms of order $(\eu\ed)^3$. It is quite natural to expect that this pattern 
extends also to higher orders.

Our detailed analysis shows that when both deformation parameters $\eu$ and $\ed$ are non-vanishing,
besides the $\epsilon$-dependent structures generated by the 
Nekrasov partition functions, the effective theory seems to require a new series of 
explicit $\eu\ed$-corrections in order to have S-duality acting in the proper way.
These new corrections, being proportional to inverse powers of the coupling constant, appear
to correspond to perturbative terms at higher loops and are absent in the Nekrasov-Shatashvili
limit. We find remarkable that by using the modified prepotential $\widehat F$ and the modified 
coupling $\widehat\tau$, all S-duality relations acquire the standard simple form 
as in the undeformed theory (see (\ref{SaFhat}), (\ref{stauhat}) and (\ref{legendre2})).
It would be very nice to see whether these results admit an interpretation in the context
of special geometry or in a more general geometric set-up 
that allows to go beyond the perturbative approach in the deformation parameters we have used 
in this paper. It would be interesting also to study the recursion relations obeyed by the functions 
$\varphi_\ell$'s which can be iteratively obtained from the viscous Burgers equation 
satisfied by $\varphi_1$ (see (\ref{Dphi1})).

We conclude by observing that all S-duality formulas we have derived for the SU(2) $N_f=4$ theory 
can be obtained for the $\cN=2^*$ SU(2) theory as well. The only difference in this case 
is that every explicit occurrence of $\eu\ed$ has to be replaced by $(\eu\ed)/2$ to take into account 
the different normalization of the viscous term in the modular anomaly recursion relation 
(see (\ref{recn2*}) as compared to (\ref{recnf4})).

\vskip 2cm
\noindent {\large {\bf Acknowledgments}}
\vskip 0.2cm
We thank Francesco Fucito and Francisco Morales for very useful discussions.
This work was supported in part by the MIUR-PRIN contract 2009-KHZKRX. 

\vskip 1cm
\appendix

\section{Modular functions}
\label{secn:appa}
We collect here some useful formulas involving the modular functions we used.

\paragraph{$\theta$-functions:} The Jacobi $\theta$-functions are defined as
\begin{equation}
\label{thetadef}
 \theta\spin{a}{b}(v|\tau)=\sum_{n\in \Z} q^{(n-\frac{a}{2})^2}
\, \ee^{2\pi \ii (n-\frac{a}{2})(v-\frac{b}{2})}~,
\end{equation}
for $a,b=0,1$ and $q=\ee^{\pi\ii\tau}$. We simplify the notation by writing, as usual, 
$\theta_1\equiv\theta\spin{1}{1}$,
$\theta_2\equiv\theta\spin{1}{0}$, $\theta_3\equiv\theta\spin{0}{0}$, 
$\theta_4\equiv\theta\spin{0}{1}$.
The functions $\theta_a$, $a=2,3,4$, satisfy the ``\emph{aequatio identica satis 
abstrusa}''
\begin{equation}
\label{abstruse}
\theta_3^4 - \theta_2^4 -\theta_4^4=0~,
\end{equation}
and admit the following series expansions
\begin{equation}
\begin{aligned}
\theta_2(0|\tau)&= 2q^{\frac{1}{4}}\big(1+q^2+q^6+q^{12}+\cdots\big)~,\\ 
\theta_3(0|\tau)&= 1+2q+2q^4+2q^{9}+\cdots~,\\ 
\theta_4(0|\tau)&= 1-2q+2q^4-2q^{9}+\cdots~. 
\end{aligned}
\end{equation}
 
\paragraph{$\eta$-function:} The Dedekind $\eta$-function is defined by
\begin{equation}
\label{dede}
\eta(q)= q^\frac{1}{12} \prod_{n=1}^\infty (1-q^{2n})~.
\end{equation}

\paragraph{Eisenstein series:} The first Eisenstein series can be expressed as follows:
\begin{equation}
\label{E246}
\begin{aligned}
E_2 & = 1 - 24 \sum_{n=1}^\infty \sigma_{1}(n)\, q^{2n} 
= 1 - 24 q^2 -72 q^4 - 96 q^6 +\ldots~,\\
E_4 & = 1 + 240 \sum_{n=1}^\infty \sigma_{3}(n)\, q^{2n}
= 1 + 240 q^2 + 2160 q^4 + 6720 q^6 + \ldots~,\\
E_6 & = 1 - 504 \sum_{n=1}^\infty \sigma_{5}(n)\, q^{2n}
= 1 - 504 q^2 -16632 q^4 - 122976 q^6 + \ldots~,
\end{aligned}
\end{equation}
where $\sigma_{k}(n)$ is the sum of the $k$-th power of the divisors 
of $n$, i.e., $\sigma_k(n) = \sum_{d|n} d^k$.
The series $E_4$ and $E_6$ are related to the $\theta$-functions
in the following way
\begin{equation}
\label{eistotheta}
\begin{aligned}
E_4 & = \frac 12 \big(\theta_2^8 + \theta_3^8 + \theta_4^8\big)~,\quad
E_6  = \frac 12 \big(\theta_3^4 + \theta_4^4\big)
 \big(\theta_2^4 + \theta_3^4\big) \big(\theta_4^4 - \theta_2^4\big)~. 
\end{aligned}
\end{equation}
The series $E_2$, $E_4$ and $E_6$
are connected among themselves by logarithmic $q$-derivatives
according to
\begin{equation}
\label{Eisring}
q\partial_q E_2  = \frac 16\big(E_2^2-E_4\big)~,\quad
q\partial_q E_4  = \frac{2}{3}\big(E_4 E_2-E_6\big)~,\quad
q\partial_q E_6  = E_6 E_2-E_4^2~.
\end{equation}
Also the derivatives of the functions $\theta_a^4$ have simple expressions:
\begin{equation}
\label{dertheta}
\begin{aligned}
q\partial_q \theta_2^4 & = \frac{\theta_2^4}{3}
\big(E_2+\theta_3^4 + \theta_4^4\big)~,\quad
q\partial_q \theta_3^4  = \frac{\theta_3^4}{3}
\big(E_2+\theta_2^4 -\theta_4^4\big)~,\quad
q\partial_q \theta_4^4 & = \frac{\theta_4^4}{3}
\big(E_2-\theta_2^4 -\theta_3^4 \big)~.
\end{aligned}
\end{equation}

\paragraph{Modular transformations:}
Under the Sl$(2,\mathbb Z)$ modular transformation
\begin{equation}
\label{modular}
\tau\to \tau' = \frac{a\tau + b}{c\tau+d}\qquad\mbox{with}\quad a,b,c,d \in\mathbb{Z}\quad\mbox{and}\quad ad-bc=1~,
\end{equation}
the Eisenstein series $E_{4}$ and $E_6$ behave as modular forms
of weight $4$ and $6$, respectively:
\begin{equation}
\label{modforms}
E_4(\tau') = (c\tau + d)^4 \,E_4(\tau)~,~~~
E_6(\tau') = (c\tau + d)^6 \,E_6(\tau)~.
\end{equation}
The series $E_2$, instead, is a quasi modular form of degree 2:
\begin{equation}
\label{E2mod}
E_2(\tau') = (c\tau + d)^2 \,E_2(\tau) + \frac{6}{\ii\pi}\,c\,(c\tau + d)~. 
\end{equation}
Under the generators $\cT$ and $\cS$ of the modular group the $\theta$-functions and the Dedekind $\eta$ function transform as follows
\begin{equation}
\begin{aligned}
&\cT~:~~~~ \theta_3^4 \leftrightarrow \theta_4^4 ~,~~~ 
\theta_2^4 \to \theta_2~,~~~
\eta\to \ee^{\frac{\ii\pi}{12}} \,\eta~,\phantom{\Bigg|}\\
&\cS~:~~~~ \theta_2^4  \to -\tau^2\, \theta_4^4~,~~~
\theta_3^4  \to -\tau^2 \,\theta_3^4~,~~~
 \theta_4^4  \to -\tau^2 \,\theta_2^4~,~~~
\eta\to \sqrt{-\ii\tau}\, \eta~.
\end{aligned}
\label{modtran1}
\end{equation}

\section{The coefficient $h_3$ of the SU(2) $N_f=4$ prepotential}
\label{secn:appb}
Here we give the explicit expression of the coefficient $h_3$ up to three instantons obtained using localization. It is given by
\begin{align}
 h_3 =&~ \frac{1}{1344}\Big[
768 R^4 - 21504 R^2 T_1 + 12288 R N + 26112 T_1^2 - 6144 T_1 T_2 - 6144 T_2^2\notag\\
&~~~~~ -(1792 R^3 - 26880 R T_1 + 10752 N)(s^2 - 2p) + R^2(1568 s^4 - 6272 s^2 p + 5376 p^2)\notag\\
&~~~~~ - T_1(9408 s^4 - 37632 s^2p + 32256 p^2) - R(496 s^6 - 2976 s^4p + 5248 s^2 p^2 - 2560 p^3)\notag\\
&~~~~~+ 51 s^8 - 408 s^6p + 1102 s^4 p^2 - 1144 s^2 p^3 + 357 p^4\Big]
 + 24 s^4(T_1 + 2 T_2) q\notag\\
&~- \frac{1}{4}\Big[3840(T_1 + 2 T_2)^2 - 1152 (R T_1 + 2 N) (5s^2 - 2 p) 
\notag\\
&~~~~~+48 (R^2 + 6 T_1) (16 s^4 -32 s^2 p +7 p^2) - 192 R(2 s^6 - 8 s^4 p + 7 s^2 p^2 - p^3)
\notag\\
&~~~~~+3(16 s^8 - 96 s^6 p +168 s^4 p^2 -88 s^2 p^3 +9 p^4)\Big]q^2\notag\\
&~ + 32 (T_1 + 2 T_2) \Big[
160(R^2 + 6 T_1) - 2R(280 s^2 - 272 p) \notag\\
&~~~~~+ 313 s^4 - 856 s^2 p + 366 p^2\Big]q^3 + \cO(q^4)~.
\label{h3nf4}
\end{align}
Following the procedure described in the main text, we can rewrite $h_3$
in terms of (quasi) modular functions according to
\begin{align}
 h_3=&~\frac{1}{3456}\big(4R - s^2 +p\big)\big(4R - s^2 +3p\big)\Big[80 R^2-8R(5s^2-22p)+5s^4-44s^2
 p+99p^2\Big]\,E_2^3
\notag\\&~
-\frac{1}{12}\Big[80 R^2-8R(5s^2-22p)+5s^4-44s^2
 p+99p^2\Big]\,(T_1\theta_4^4-T_2\theta_2^4)\,E_2^2
\notag \\
&~+\frac{1}{1440}\big(4R - s^2 +6p\big)\Big[64 R^3-80 R^2(3s^2-4p)+4R(27s^4-88s^2p+49p^2)\notag\\
&~~~~~~~~~~~~-13s^6+68s^4p-94s^2p^2+30p^3+2304 N\Big]\,E_2\,E_4\notag\\
&~-\frac{4}{3}\big(4R - s^2 +6p\big)\big(R - s^2 +p\big)
\Big[T_1\theta_4^4(2\theta_2^4+\theta_4^4)+T_2\theta_2^4(\theta_2^4+2\theta_4^4)
\Big]\,E_2\notag\\
&~+8( T_1\theta_4^4- T_2\theta_2^4)^2 E_2
+16\Big[T_1^2(\theta_4^4+2\theta_2^4)\theta_4^8 -
T_2^2(\theta_2^4+2\theta_4^4)\theta_2^8 -T_1T_2\,\theta_2^4\theta_4^4(\theta_2^4-\theta_4^4) \Big]\notag\\
&~-\frac{1}{4}\Big[16 R^2-8R(5s^2-6p)+21s^4-60s^2
 p+31p^2\Big]\,(T_1\theta_4^4-T_2\theta_2^4)\,E_4\notag\\
&~+\frac{1}{120960}\Big[2816 R^4-30464R^3 s^2 +39424 R^3 p+67872 R^2 s^4-190848R^2 s^2 p\notag\\
&~~~~+94304 R^2 p^2-28400 R s^6+134112 R s^4 p-165664 R s^2 p^2+47616 R p^3\notag \\
&~~~~-774144N(s^2-p)+331776 N R-552960(T_1^2+T_1T_2+T_2^2)\notag\\
&~~~~+3323 s^8-22216 s^6 p+46862 s^4 p^2-34584 s^2p^3+6615 p^4\Big] E_6~.
\end{align}
By expanding the modular functions in powers of $q$ as shown in Appendix \ref{secn:appa}, one can
recover the instanton terms in (\ref{h3nf4}).

\section{Reformulating the modular anomaly equations}
\label{secn:appc}
The modular anomaly equation (\ref{recnf4}) implies the following relation:
\begin{equation}
D \varphi_0 = \frac{1}{2}\,\varphi_1^2+\frac{\eu\ed}{2}\, \varphi_2
\label{Dphi0}
\end{equation}
where the functions $\varphi_\ell$'s have been defined in (\ref{phil}).
Since the operators $D$ and $(-\partial_a)$ commute, by applying the latter to (\ref{Dphi0})
and remembering that $\varphi_{\ell+1}=-\partial_a\varphi_\ell$,
it is straightforward to obtain the action of $D$ on any
$\varphi_\ell$ and verify that these functions form a ring under it.
For example, at the next step we obtain
\begin{equation}
 D \varphi_1 = \varphi_1\varphi_2 + \frac{\eu\ed}{2}\, \varphi_3  =
\varphi_1 \big(\!-\partial_a\varphi_1\big) + \frac{\eu\ed}{2}\,\partial_a^2\varphi_1~.
  \label{Dphi1}
\end{equation}
which is, up to rescalings, the viscous Burgers equation
\begin{equation}
\label{Burgers}
\partial_t u = u \partial_x u + \nu \partial^2_x u
\end{equation}
with the viscosity $\nu$ proportional to $\eu\ed$.

In Section~\ref{secn:sdual} the S-duality requirements were formulated 
in terms of the quantities $\Sigma^{(\ell)}$ defined in (\ref{sigmas}), which we rewrite here for convenience as follows:
\begin{equation}
\label{defSigmaP}
\Sigma^{(\ell)} = \sum_{n=0}^\infty\frac{1}{n!\,(2\pi\ii\tau_0)^n}\, P_n^{(\ell)}
\end{equation}
with
\begin{equation}
\label{defP}
P_n^{(\ell)} = \left(D+\zeta\,\partial_a\right)^n\varphi_\ell\,\Big|_{\zeta=\varphi_1}~.
\end{equation}
{From} this definition it is easy to show that the following relation holds:
\begin{equation}
\label{newrecP}
P_n^{(\ell+1)}=-\partial_a P_n^{(\ell)}  + n\,\varphi_2 P_{n-1}^{(\ell+1)}~.
\end{equation}
In turn, this implies that 
\begin{equation}
\Sigma^{(\ell+1)} = - \partial_a \Sigma^{(\ell)} + \frac{\varphi_2}{2\pi\ii\tau_0}\Sigma^{(\ell+1)}~.
\end{equation}
Recalling that $2\pi\ii\tau_0 - \varphi_2= 2\pi\ii\tau$ (see (\ref{phi2F})), this is tantamount
to the recursion relation (\ref{sigmal}), namely
\begin{equation}
 \Sigma^{(\ell+1)} = -\frac{\tau_0}{\tau}\,\partial_a \Sigma^{(\ell)}~,
\label{sigmalreplay}
\end{equation}
which allows to easily obtain the expression of any $\Sigma^{(\ell)}$ once $\Sigma^{(0)}$ 
is known.

To determine $\Sigma^{(0)}$ we start considering the quantities $P^{(0)}_n$.
Clearly, from (\ref{defP}) we have $P_0^{(0)}=\varphi_0$. The next case is  
\begin{equation}
\label{P10}
\begin{aligned}
P_1^{(0)} &= \big(D+\zeta\,\partial_a\big)\varphi_0\Big|_{\zeta=\varphi_1} 
= D\varphi_0-\varphi_1^2= -\frac{1}{2}\varphi_1^2+\frac{\eu\ed}{2}\,\varphi_2
\end{aligned}
\end{equation}
where the last step follows from (\ref{Dphi0}). By further applications of the operator $\big(D+\zeta\,\partial_a\big)$, we get
\begin{equation}
\label{firstPs}
\begin{aligned}
P_2^{(0)} &=\frac{\epsilon_1\epsilon_2}{2}\,\varphi_2^2 +\left(\frac{\epsilon_1\epsilon_2}{2} \right)^2 \varphi_4~,\\
P_3^{(0)} &=\epsilon_1\epsilon_2 \,\varphi_2^3 +\left(\frac{\epsilon_1\epsilon_2}{2} \right)^2
\big(6\varphi_2 \varphi_4+5\varphi_3^2\big)
+\left(\frac{\epsilon_1\epsilon_2}{2} \right)^3\varphi_6~,
\end{aligned}
\end{equation}
with similar expressions for higher values of $n$. In fact, it is possible to derive the general expression of the $P_n^{(0)}$'s at the first orders in their explicit dependence on $\eu\ed$:
\begin{equation}
\label{P0eexp}
\begin{aligned}
 P_n^{(0)}\Big|_0 & = \delta_{n,0}\,\varphi_0 -\delta_{n,1}\,\frac{\varphi_1^2}{2}~,\\
 P_n^{(0)}\Big|_1 & = \frac{(n-1)!}{2}\,\varphi_2^{n}~,\\
 P_n^{(0)}\Big|_2 & = \frac{(n-1)!}{4}\, 
\Big(\frac 12 \varphi_4 \pffb^2 + \frac{5}{12} \varphi_3^2 \pffb^3 \Big) \varphi_2^n~.
\end{aligned}
\end{equation}
Inserting these results into (\ref{defSigmaP}), we obtain 
\begin{equation}
 \begin{aligned}
\Sigma^{(0)}\Big|_{0} &= \varphi_0-\frac{\varphi_1^2}{2(2\pi\ii\tau_0)}~,\\
\Sigma^{(0)}\Big|_{1} &=-\frac{1}{2} \log \Big(1- \frac{\varphi_2}{2\pi\ii\tau_0}\Big)
= -\frac{1}{2} \log \frac{\tau}{\tau_0}
~,\\
\Sigma^{(0)}\Big|_{2} &=
\frac{1}{8}
\frac{\varphi_4}{(2\pi\ii\tau)^2}+\frac{5}{24} \frac{\varphi_3^2}{(2\pi\ii\tau)^3}~,
 \end{aligned}
\label{sumPn0}
\end{equation}
where from the second equality on we have used that
$2\pi\ii\tau_0 - \varphi_2= 2\pi\ii\tau$. 

Applying the recursion formula (\ref{sigmalreplay})
we can easily get the explicit expressions for the $\Sigma^{(\ell)}$'s that are needed in the
calculations presented in Section~\ref{secn:sdual} or in Appendix~\ref{secn:appd}. 
They are, for $\ell=1$:
\begin{equation}
 \begin{aligned}
\Sigma^{(1)}\Big|_{0} &= \varphi_1~,\\
\Sigma^{(1)}\Big|_{1} &=2\pi\ii\tau_0\Big(\frac{1}{2}\,\frac{\varphi_3}{(2\pi\ii\tau)^2}\Big)~,
\\
\Sigma^{(1)}\Big|_{2} &=2\pi\ii\tau_0\Big(\frac{1}{8}\frac{\varphi_5}{(2\pi\ii\tau)^3}
+\frac{2}{3}\frac{\varphi_3\varphi_4}{(2\pi\ii\tau)^4}+
\frac{5}{8}\frac{\varphi_3^3}{(2\pi\ii\tau)^5}\Big)~;
 \end{aligned}
\label{sumPn1}
\end{equation}
for $\ell=2$:
\begin{equation}
 \begin{aligned}
\Sigma^{(2)}\Big|_{0} &= 2\pi\ii\tau_0\Big(\frac{\varphi_2}{2\pi\ii\tau}\Big)~,\\
\Sigma^{(2)}\Big|_{1} &=
(2\pi\ii\tau_0)^2\Big(\frac{1}{2}\,\frac{\varphi_4}{(2\pi\ii\tau)^3} + 
\frac{\varphi_3^2}{(2\pi\ii\tau)^4}\Big)~;
 \end{aligned}
\label{sumPn2}
\end{equation}
for $\ell=3$:
\begin{equation}
 \begin{aligned}
\Sigma^{(3)}\Big|_{0} &= (2\pi\ii\tau_0)^3\Big(\frac{\varphi_3}{(2\pi\ii\tau)^3}\Big)~,\\
\Sigma^{(3)}\Big|_{1} &= 
(2\pi\ii\tau_0)^3\Big(\frac{1}{2}\,\frac{\varphi_5}{(2\pi\ii\tau)^4}
+\frac{7}{2}\,\frac{\varphi_3\varphi_4}{(2\pi\ii\tau)^5}
+4\,\frac{\varphi_3^3}{(2\pi\ii\tau)^6}\Big)~;
 \end{aligned}
\label{sumPn3}
\end{equation}
for $\ell=4:$
\begin{equation}
 \Sigma^{(4)}\Big|_{0} =
(2\pi\ii\tau_0)^4\Big(\frac{\varphi_4}{(2\pi\ii\tau)^4}+3\,\frac{\varphi_3^2}{(2\pi\ii\tau)^5}\Big)~;
\label{sumPn4}
\end{equation}
and finally for $\ell=5:$
\begin{equation}
 \begin{aligned}
\Sigma^{(5)}\Big|_{0} &= (2\pi\ii\tau_0)^5\Big(
\frac{\varphi_5}{(2\pi\ii\tau)^5}+10\,\frac{\varphi_3\varphi_4}{(2\pi\ii\tau)^6}
+15\,\frac{\varphi_3^3}{(2\pi\ii\tau)^7}\Big)~.
 \end{aligned}
\label{sumPn5}
\end{equation}

\section{S-duality at order $(\eu\ed)^2$}
\label{secn:appd}
We start from the relation (\ref{Sphi11}) for $n=2$, namely
\begin{equation}
\label{sdn2}
\tau_0\, \cS(X_2)\Big|_0 - X_2 = - \tau_0 S(\varphi_1)\Big|_2 - \tau_0 S(X_1)\Big|_1~. 
\end{equation}
{From} (\ref{tSphilX}) we read that
\begin{equation}
\label{sphi1at2}
\tau_0 \,\cS(\varphi_1)\Big|_2  = \Sigma^{(1)}\Big|_2 - \frac{X_1}{2\pi\ii\tau_0} \Sigma^{(2)}\Big|_1 - 
\frac{X_2}{2\pi\ii\tau_0} \Sigma^{(2)}\Big|_0 + \frac 12 \frac{X_1^2}{(2\pi\ii\tau_0)^2}
\Sigma^{(3)}\Big|_0~.
\end{equation}
Substituting the expressions of the various $\Sigma^{(\ell)}\Big|_k$'s
given in Appendix~\ref{secn:appc}, and that of $X_1$ given in (\ref{X1fin}), we get 
\begin{equation}
\label{sphi1at2bis}
\tau_0 \,\cS(\varphi_1)\Big|_2  = 2\pi\ii\tau_0\Big(\frac{1}{8}\,\frac{\varphi_5}{(2\pi\ii\tau)^3}
+\frac{13}{24}\,\frac{\varphi_4\varphi_3}{(2\pi\ii\tau)^4}
+\frac{13}{32}\,\frac{\varphi_3^3}{(2\pi\ii\tau)^5}\Big)
-\frac{\varphi_2}{2\pi\ii\tau}X_2~.
\end{equation}
On the other hand, from (\ref{X1fin}) it follows that
\begin{equation}
\label{SX1d}
\tau_0 \,\cS(X_1) = \frac{1}{8\pi\ii\tau_0^2} 
\frac{\tau_0^3 \,\cS(\varphi_3)}{S(\tau)}~.
\end{equation}
The numerator of this expression can be computed from (\ref{tSphilX}), yielding
\begin{eqnarray}
 \tau_0^3\, \cS(\varphi_3) &=& \Sigma^{(3)} - \frac{X}{2\pi\ii\tau_0} \Sigma^{(4)} + \cdots
= \Sigma^{(3)}\Big|_0  + \eu\ed\Big(\Sigma^{(3)}\Big|_1 - \frac{X_1}{2\pi\ii\tau_0} \Sigma^{(4)}\Big|_0 
\Big)+ \cdots
\label{Sphi3}\\
&=&(2\pi\ii\tau_0)^3\,\frac{\varphi_3}{(2\pi\ii\tau)^3}
+\eu\ed\,(2\pi\ii\tau_0)^3\Big(\frac{1}{2}\,\frac{\varphi_5}{(2\pi\ii\tau)^4}
+\frac{13}{4}\,\frac{\varphi_3\varphi_4}{(2\pi\ii\tau)^5}
+\frac{13}{4}\,\frac{\varphi_3^3}{(2\pi\ii\tau)^6}
\Big)+\cdots
\nonumber
\end{eqnarray}
where the last step follows from the results given in Appendix~\ref{secn:appc}.
For the denominator we take into account that 
\begin{equation}
\label{Stau1d}
\cS(\tau) = \cS(\tau_0)-\frac{\cS(\varphi_2)}{2\pi\ii} = - \frac{1}{\tau_0} - \frac{\cS(\varphi_2)}{2\pi\ii}~.
\end{equation}
Resorting again to (\ref{tSphilX}) to evaluate $\tau_0^2\,\cS(\varphi_2)$, we obtain
\begin{equation}
\label{Stau2}
\begin{aligned}
\cS(\tau) & = -\frac{1}{\tau_0} -\frac{1}{2\pi\ii\tau_0^2}\Bigg[\Sigma^{(2)}\Big|_0 + 
\eu\ed \Big(\Sigma^{(2)}\Big|_1 - \frac{X_1}{2\pi\ii\tau_0}\Sigma^{(3)}\Big|_0\Big)+ \cdots\Bigg]\\
&=-\frac{1}{\tau}-\eu\ed\Big(\frac{1}{2}\,\frac{\varphi_4}{(2\pi\ii)^2\tau^3}+
\frac{3}{4}\,\frac{\varphi_3^2}{(2\pi\ii)^3\tau^4}\Big)+\cdots~.
\end{aligned}
\end{equation}
Inserting (\ref{Sphi3}) and (\ref{Stau2}) into (\ref{SX1d}) and extracting the term of 
order $\eu\ed$, we find
\begin{equation}
\label{SX1tris}
\begin{aligned}
\tau_0\,\cS(X_1)\Big|_1 & = -(2\pi\ii\tau_0)\Big(
\frac{1}{8}\,\frac{\varphi_5}{(2\pi\ii\tau)^3}
+\frac{11}{16}\,\frac{\varphi_3\varphi_4}{(2\pi\ii\tau)^4}
+\frac{5}{8}\,\frac{\varphi_3^3}{(2\pi\ii\tau)^5}\Big)~.
\end{aligned}
\end{equation}
Using this result and (\ref{sphi1at2bis}) into (\ref{sdn2}), we finally obtain the following constraint on $X_2$:
\begin{equation}
\label{conX2}
\tau_0\,\cS(X_2)\Big|_0 - 2\pi\ii\tau_0\Big(\frac{X_2}{2\pi\ii\tau}\Big) = (2\pi\ii\tau_0)\Big(\frac{7}{48}\,\frac{\varphi_3\varphi_4}{(2\pi\ii\tau)^4}
+\frac{7}{32}\,\frac{\varphi_3^3}{(2\pi\ii\tau)^5}\Big)~.
\end{equation}
Considering the structures involved in this relation, we can try an \emph{Ansatz} such that $X_2$ is written as a total derivative:
\begin{equation}
\label{X2der}
\begin{aligned}
 X_2 &= \partial_a\left(-\lambda_1 \frac{\varphi_4}{(2\pi\ii\tau)^2} - \lambda_2
\frac{\varphi_3^2}{(2\pi\ii\tau)^3}\right) \\
&=\lambda_1 \frac{\varphi_5}{(2\pi\ii\tau)^2} + 2(\lambda_1+\lambda_2) \frac{\varphi_3\varphi_4}{(2\pi\ii\tau)^3}
+ 3\lambda_2 \frac{\varphi_3^3}{(2\pi\ii\tau)^4}~.
\end{aligned}
\end{equation}
With this position we find straightforwardly, using the formul\ae\,
of Appendix~\ref{secn:appc}, that 
\begin{equation}
\label{SX2a}
\tau_0\,\cS(X_2)\Big|_0 - 2\pi\ii\tau_0\Big(\frac{X_2}{2\pi\ii\tau}\Big) =2\pi\ii\tau_0\Big( 
(6\lambda_1- 4\lambda_2) \,\frac{\varphi_3\varphi_4}{(2\pi\ii\tau)^4}
+(9\lambda_1 - 6 \lambda_2 )\,\frac{\varphi_3^3}{(2\pi\ii\tau)^5}\Big)~.
\end{equation}
Comparing this result to the constraint (\ref{conX2}), we obtain a single independent equation:
\begin{equation}
\label{eqpar1}
6\lambda_1 - 4\lambda_2 = \frac{7}{48}~.
\end{equation}
Our \emph{Ansatz} thus satisfies the requirement (\ref{conX2}) with still one free parameter. 
In Section~\ref{secn:conl} we chose to fix this arbitrariness by requiring that 
the modified prepotential
$\widehat F$ introduced in (\ref{SaFhat}) differs from $F$ by a series of corrections in powers of $\eu\ed$ whose coefficients only involve $\varphi_3$ and the modified effective 
coupling $\widehat\tau$ defined in (\ref{tauhat}). It is easy to see that, starting with 
$X_2$ given in (\ref{X2der}), the $\widehat F$ term of order $(\eu\ed)^2$
in (\ref{Fhat1}) would contain a part proportional to $\varphi_4$ unless 
$\lambda_1$ assumes a specific value which, by virtue of (\ref{eqpar1}), fixes also $\lambda_2$.
These values are:
\begin{equation}
\label{lambdachoice}
\lambda_1 = \frac{1}{16}~,~~~
\lambda_2 = \frac{11}{192}~,
\end{equation} 
which are those used in (\ref{X2}).

\providecommand{\href}[2]{#2}\begingroup\raggedright
\endgroup

\end{document}